
1 Postscript figure is found at XXX.

%
%
  \font\twelverm=cmr10 scaled 1200       \font\twelvei=cmmi10 scaled 1200
  \font\twelvesy=cmsy10 scaled 1200      \font\twelveex=cmex10 scaled 1200
  \font\twelvebf=cmbx10 scaled 1200      \font\twelvesl=cmsl10 scaled 1200
  \font\twelvett=cmtt10 scaled 1200      \font\twelveit=cmti10 scaled 1200
  \font\twelvemib=cmmib10 scaled 1200
  \font\elevenmib=cmmib10 scaled 1095
  \font\tenmib=cmmib10
  \font\eightmib=cmmib10 scaled 800
  
\font\elevenrm=cmr10 scaled 1095    \font\eleveni=cmmi10 scaled 1095
\font\elevensy=cmsy10 scaled 1095

%
%

\font\seventeeni=cmmi10 scaled \magstep3

\font\seventeensy=cmsy10 scaled \magstep3

\font\seventeenmib=cmmib10 scaled \magstep3

\newfam\cpfam%



\skewchar\eleveni='177   \skewchar\elevensy='60
\skewchar\elevenmib='177  \skewchar\seventeensy='60
\skewchar\seventeenmib='177
\skewchar\seventeeni='177

\newfam\mibfam%


  \skewchar\twelvei='177   \skewchar\twelvesy='60
  \skewchar\twelvemib='177
%
%
\def\twelvepoint{\normalbaselineskip=12.4pt
  \abovedisplayskip 12.4pt plus 3pt minus 9pt
  \belowdisplayskip 12.4pt plus 3pt minus 9pt
  \abovedisplayshortskip 0pt plus 3pt
  \belowdisplayshortskip 7.2pt plus 3pt minus 4pt
  \smallskipamount=3.6pt plus 1.2pt minus 1.2pt
  \medskipamount=7.2pt plus 2.4pt minus 2.4pt
  \bigskipamount=14.4pt plus 4.8pt minus 4.8pt
  \def\rm{\fam0\twelverm}          \def\it{\fam\itfam\twelveit}%
  \def\sl{\fam\slfam\twelvesl}     \def\bf{\fam\bffam\twelvebf}%
  \def\mit{\fam 1}                 \def\cal{\fam 2}%
  \def\tt{\twelvett}%
  \def\mib{\fam\mibfam\twelvemib}%

  \textfont0=\twelverm   \scriptfont0=\tenrm     \scriptscriptfont0=\sevenrm
  \textfont1=\twelvei    \scriptfont1=\teni      \scriptscriptfont1=\seveni
  \textfont2=\twelvesy   \scriptfont2=\tensy     \scriptscriptfont2=\sevensy
  \textfont3=\twelveex   \scriptfont3=\twelveex  \scriptscriptfont3=\twelveex
  \textfont\itfam=\twelveit
  \textfont\slfam=\twelvesl
  \textfont\bffam=\twelvebf
  \textfont\mibfam=\twelvemib       \scriptfont\mibfam=\tenmib
                                             \scriptscriptfont\mibfam=\eightmib

  \def\xrm{\textfont0=\twelverm\scriptfont0=\tenrm
      \scriptscriptfont0=\sevenrm\rm}
\normalbaselines\rm}


\mathchardef\alpha="710B
\mathchardef\beta="710C
\mathchardef\gamma="710D
\mathchardef\delta="710E
\mathchardef\epsilon="710F
\mathchardef\zeta="7110
\mathchardef\eta="7111
\mathchardef\theta="7112
\mathchardef\kappa="7114
\mathchardef\lambda="7115
\mathchardef\mu="7116
\mathchardef\nu="7117
\mathchardef\xi="7118
\mathchardef\pi="7119
\mathchardef\rho="711A
\mathchardef\sigma="711B
\mathchardef\tau="711C
\mathchardef\phi="711E
\mathchardef\chi="711F
\mathchardef\psi="7120
\mathchardef\omega="7121
\mathchardef\varepsilon="7122
\mathchardef\vartheta="7123
\mathchardef\varrho="7125
\mathchardef\varphi="7127

\def\physgreek{
\mathchardef\Gamma="7100
\mathchardef\Delta="7101
\mathchardef\Theta="7102
\mathchardef\Lambda="7103
\mathchardef\Xi="7104
\mathchardef\Pi="7105
\mathchardef\Sigma="7106
\mathchardef\Upsilon="7107
\mathchardef\Phi="7108
\mathchardef\Psi="7109
\mathchardef\Omega="710A}


\def\beginlinemode{\endmode
  \begingroup\parskip=0pt \obeylines\def\\{\par}\def\endmode{\par\endgroup}}
\def\beginparmode{\endmode
  \begingroup \def\endmode{\par\endgroup}}
\let\endmode=\par
{\obeylines\gdef\
{}}
\def\singlespace{\baselineskip=\normalbaselineskip}

\def\oneandahalfspace{\baselineskip=\normalbaselineskip
  \multiply\baselineskip by 3 \divide\baselineskip by 2}
\def\doublespace{\baselineskip=\normalbaselineskip \multiply\baselineskip by 2}

\nopagenumbers
\newcount\firstpageno
\firstpageno=2
\headline={\ifnum\pageno<\firstpageno{\hfil}\else{\hfil\elevenrm\folio\hfil}\fi}
\let\rawfootnote=\footnote             
\def\footnote#1#2{{\singlespace\parindent=0pt
\rawfootnote{#1}{#2}}}
\def\raggedcenter{\leftskip=4em plus 12em \rightskip=\leftskip
  \parindent=0pt \parfillskip=0pt \spaceskip=.3333em \xspaceskip=.5em
  \pretolerance=9999 \tolerance=9999
  \hyphenpenalty=9999 \exhyphenpenalty=9999 }
\def\dateline{\rightline{\ifcase\month\or
  January\or February\or March\or April\or May\or June\or
  July\or August\or September\or October\or November\or December\fi
  \space\number\year}}
\def\received{\vskip 3pt plus 0.2fill
 \centerline{\sl (Received\space\ifcase\month\or
  January\or February\or March\or April\or May\or June\or
  July\or August\or September\or October\or November\or December\fi
  \qquad, \number\year)}}


\hsize=6.5truein
\hoffset=0.0truein
\vsize=8.9truein
\voffset=0truein
\hfuzz=0.1pt
\vfuzz=0.1pt
\parskip=\medskipamount
\overfullrule=0pt      



\def\title                     
  {\null\vskip 3pt plus 0.1fill
   \beginlinemode \doublespace \raggedcenter \bf}

\def\author                    
  {\vskip 6pt plus 0.2fill \beginlinemode
   \singlespace \raggedcenter}

\def\affil        
  {\vskip 6pt plus 0.1fill \beginlinemode
   \oneandahalfspace \raggedcenter \it}

\def\abstract                  
  {\vskip 6pt plus 0.3fill \beginparmode
   \doublespace \narrower }

\def\summary                   
  {\vskip 3pt plus 0.3fill \beginparmode
   \doublespace \narrower SUMMARY: }

\def\pacs#1
  {\vskip 3pt plus 0.2fill PACS numbers: #1}

\def\endtitlepage              
  {\endpage                    
   \body}

\def\body                      
  {\beginparmode}              

\def\head#1{                   
  \filbreak\vskip 0.5truein    
  {\immediate\write16{#1}
   \raggedcenter \uppercase{#1}\par}
   \nobreak\vskip 0.25truein\nobreak}

%
%

%
\def\inlinerefs{
  \gdef\refto##1{ [##1]}                
\gdef\refis##1{\indent\hbox to 0pt{\hss##1.~}} 
\gdef\journal##1, ##2, ##3, 1##4##5##6{ 
    {\sl ##1~}{\bf ##2}, ##3 (1##4##5##6)}}    
\def\keywords#1
  {\vskip 3pt plus 0.2fill Keywords: #1}
\gdef\figis#1{\indent\hbox to 0pt{\hss#1.~}} 

\def\figurecaptions     
  {\head{Figure Captions}    
   \beginparmode
   \interlinepenalty=10000
   \frenchspacing \parindent=0pt \leftskip=1truecm
   \parskip=8pt plus 3pt \everypar{\hangindent=\parindent}}

%
%
\def\refto#1{$^{#1}$}          

\def\references       
  {\head{References}           
   \beginparmode
   \frenchspacing \parindent=0pt \leftskip=1truecm
   \interlinepenalty=10000
   \parskip=8pt plus 3pt \everypar{\hangindent=\parindent}}

\gdef\refis#1{\indent\hbox to 0pt{\hss#1.~}} 

\gdef\journal#1, #2, #3, 1#4#5#6{              
    {\sl #1~}{\bf #2}, #3 (1#4#5#6)}          

\def\refstylenp{               
  \gdef\refto##1{ [##1]}                               
  \gdef\refis##1{\indent\hbox to 0pt{\hss##1)~}}      
  \gdef\journal##1, ##2, ##3, ##4 {                    
     {\sl ##1~}{\bf ##2~}(##3) ##4 }}

\def\refstyleprnp{             
  \gdef\refto##1{ [##1]}                               
  \gdef\refis##1{\indent\hbox to 0pt{\hss##1)~}}      
  \gdef\journal##1, ##2, ##3, 1##4##5##6{              
    {\sl ##1~}{\bf ##2~}(1##4##5##6) ##3}}

\def\pra{\journal Phys. Rev. A, }

\def\prb{\journal Phys. Rev. B, }

\def\prl{\journal Phys. Rev. Lett., }

\def\endreferences{\body}

%
%

\def\endfigurecaptions{\body}

\def\endpage                   
  {\vfill\eject}

\def\endpaper                  
  {\endmode\vfill\supereject}

\def\endit
  {\endpaper\end}


\def\ref#1{Ref.[#1]}                   
\def\Ref#1{Ref.[#1]}                   

\def\Equation#1{Equation [#1]}         
\def\Equations#1{Equations [#1]}       
\def\Eq#1{Eq. (#1)}                     
\def\eq#1{Eq. (#1)}                     
\def\Eqs#1{Eqs. (#1)}                   
\def\eqs#1{Eqs. (#1)}                   
\def\frac#1#2{{\textstyle{{\strut #1} \over{\strut #2}}}}

\def\sla{\raise.15ex\hbox{$/$}\kern-.57em}
\def\leaderfill{\leaders\hbox to 1em{\hss.\hss}\hfill}
\def\twiddle{\lower.9ex\rlap{$\kern-.1em\scriptstyle\sim$}}
\def\bigtwiddle{\lower1.ex\rlap{$\sim$}}
\def\gtwid{\mathrel{\raise.3ex\hbox{$>$\kern-.75em\lower1ex\hbox{$\sim$}}}}
\def\ltwid{\mathrel{\raise.3ex\hbox{$<$\kern-.75em\lower1ex\hbox{$\sim$}}}}
\def\square{\kern1pt\vbox{\hrule height 1.2pt\hbox{\vrule width 1.2pt\hskip 3pt
   \vbox{\vskip 6pt}\hskip 3pt\vrule width 0.6pt}\hrule height 0.6pt}\kern1pt}

%

%

%

%
\physgreek
%

\def\dsl{\raise.15ex\hbox{$/$}\kern-.57em\hbox{$\partial$}}
\def\nsl{\raise.15ex\hbox{$/$}\kern-.57em\hbox{$\nabla$}}
\def\gtwid{\,{\raise.3ex\hbox{$>$\kern-.75em\lower1ex\hbox{$\sim$}}}\,}
\def\ltwid{\,{\raise.3ex\hbox{$<$\kern-.75em\lower1ex\hbox{$\sim$}}}\,}
\def\undr{\raise.3ex\hbox{$\sim$\kern-.75em\lower1ex\hbox{$|\vec
x|\to\infty$}}}

\def\[{\left [}
\def\]{\right ]}
\def\({\left (}
\def\){\right )}







\def\and{a^{\phantom\dagger}}

%
\def\id{\raise.72ex\hbox{$-$}\kern-.85em\hbox{$d$}\,}

\catcode`@=11
\newcount\r@fcount \r@fcount=0
\newcount\r@fcurr
\immediate\newwrite\reffile
\newif\ifr@ffile\r@ffilefalse
\def\w@rnwrite#1{\ifr@ffile\immediate\write\reffile{#1}\fi\message{#1}}

\def\writer@f#1>>{}
\def\referencefile{
  \r@ffiletrue\immediate\openout\reffile=\jobname.ref%
  \def\writer@f##1>>{\ifr@ffile\immediate\write\reffile%
    {\noexpand\refis{##1} = \csname r@fnum##1\endcsname = %
     \expandafter\expandafter\expandafter\strip@t\expandafter%
     \meaning\csname r@ftext\csname r@fnum##1\endcsname\endcsname}\fi}%
  \def\strip@t##1>>{}}

\def\citeall#1{\xdef#1##1{#1{\noexpand\cite{##1}}}}
\def\cite#1{\each@rg\citer@nge{#1}}	

\def\each@rg#1#2{{\let\thecsname=#1\expandafter\first@rg#2,\end,}}
\def\first@rg#1,{\thecsname{#1}\apply@rg}	
\def\apply@rg#1,{\ifx\end#1\let\next=\relax
\else,\thecsname{#1}\let\next=\apply@rg\fi\next}

\def\citer@nge#1{\citedor@nge#1-\end-}	
\def\citer@ngeat#1\end-{#1}
\def\citedor@nge#1-#2-{\ifx\end#2\r@featspace#1 
  \else\citel@@p{#1}{#2}\citer@ngeat\fi}	
\def\citel@@p#1#2{\ifnum#1>#2{\errmessage{Reference range #1-#2\space is bad.}%
    \errhelp{If you cite a series of references by the notation M-N, then M and
    N must be integers, and N must be greater than or equal to M.}}\else%
 {\count0=#1\count1=#2\advance\count1
by1\relax\expandafter\r@fcite\the\count0,%
  \loop\advance\count0 by1\relax
    \ifnum\count0<\count1,\expandafter\r@fcite\the\count0,%
  \repeat}\fi}

\def\r@featspace#1#2 {\r@fcite#1#2,}	
\def\r@fcite#1,{\ifuncit@d{#1}
    \newr@f{#1}%
    \expandafter\gdef\csname r@ftext\number\r@fcount\endcsname%
                     {\message{Reference #1 to be supplied.}%
                      \writer@f#1>>#1 to be supplied.\par}%
 \fi%
 \csname r@fnum#1\endcsname}
\def\ifuncit@d#1{\expandafter\ifx\csname r@fnum#1\endcsname\relax}%
\def\newr@f#1{\global\advance\r@fcount by1%
    \expandafter\xdef\csname r@fnum#1\endcsname{\number\r@fcount}}

\let\r@fis=\refis			
\def\refis#1#2#3\par{\ifuncit@d{#1}
   \newr@f{#1}%
   \w@rnwrite{Reference #1=\number\r@fcount\space is not cited up to now.}\fi%
  \expandafter\gdef\csname r@ftext\csname r@fnum#1\endcsname\endcsname%
  {\writer@f#1>>#2#3\par}}

\def\ignoreuncited{
   \def\refis##1##2##3\par{\ifuncit@d{##1}%
     \else\expandafter\gdef\csname r@ftext\csname
r@fnum##1\endcsname\endcsname%
     {\writer@f##1>>##2##3\par}\fi}}

\def\r@ferr{\endreferences\errmessage{I was expecting to see
\noexpand\endreferences before now;  I have inserted it here.}}
\let\r@ferences=\references
\def\references{\r@ferences\def\endmode{\r@ferr\par\endgroup}}

\let\endr@ferences=\endreferences
\def\endreferences{\r@fcurr=0
  {\loop\ifnum\r@fcurr<\r@fcount
    \advance\r@fcurr by 1\relax\expandafter\r@fis\expandafter{\number\r@fcurr}%
    \csname r@ftext\number\r@fcurr\endcsname%
  \repeat}\gdef\r@ferr{}\endr@ferences}


\let\r@fend=\endpaper\gdef\endpaper{\ifr@ffile
\immediate\write16{Cross References written on []\jobname.REF.}\fi\r@fend}

\catcode`@=12

\citeall\refto		
\citeall\ref		%
\citeall\Ref		%

\catcode`@=11
\newcount\tagnumber\tagnumber=0

\immediate\newwrite\eqnfile
\newif\if@qnfile\@qnfilefalse
\def\write@qn#1{}
\def\writenew@qn#1{}
\def\w@rnwrite#1{\write@qn{#1}\message{#1}}
\def\@rrwrite#1{\write@qn{#1}\errmessage{#1}}

\def\taghead#1{\gdef\t@ghead{#1}\global\tagnumber=0}
\def\t@ghead{}

\expandafter\def\csname @qnnum-3\endcsname
  {{\t@ghead\advance\tagnumber by -3\relax\number\tagnumber}}
\expandafter\def\csname @qnnum-2\endcsname
  {{\t@ghead\advance\tagnumber by -2\relax\number\tagnumber}}
\expandafter\def\csname @qnnum-1\endcsname
  {{\t@ghead\advance\tagnumber by -1\relax\number\tagnumber}}
\expandafter\def\csname @qnnum0\endcsname
  {\t@ghead\number\tagnumber}
\expandafter\def\csname @qnnum+1\endcsname
  {{\t@ghead\advance\tagnumber by 1\relax\number\tagnumber}}
\expandafter\def\csname @qnnum+2\endcsname
  {{\t@ghead\advance\tagnumber by 2\relax\number\tagnumber}}
\expandafter\def\csname @qnnum+3\endcsname
  {{\t@ghead\advance\tagnumber by 3\relax\number\tagnumber}}

\def\equationfile{%
  \@qnfiletrue\immediate\openout\eqnfile=\jobname.eqn%
  \def\write@qn##1{\if@qnfile\immediate\write\eqnfile{##1}\fi}
  \def\writenew@qn##1{\if@qnfile\immediate\write\eqnfile
    {\noexpand\tag{##1} = (\t@ghead\number\tagnumber)}\fi}
}

\def\callall#1{\xdef#1##1{#1{\noexpand\call{##1}}}}
\def\call#1{\each@rg\callr@nge{#1}}

\def\each@rg#1#2{{\let\thecsname=#1\expandafter\first@rg#2,\end,}}
\def\first@rg#1,{\thecsname{#1}\apply@rg}
\def\apply@rg#1,{\ifx\end#1\let\next=\relax%
\else,\thecsname{#1}\let\next=\apply@rg\fi\next}

\def\callr@nge#1{\calldor@nge#1-\end-}
\def\callr@ngeat#1\end-{#1}
\def\calldor@nge#1-#2-{\ifx\end#2\@qneatspace#1 %
  \else\calll@@p{#1}{#2}\callr@ngeat\fi}
\def\calll@@p#1#2{\ifnum#1>#2{\@rrwrite{Equation range #1-#2\space is bad.}
\errhelp{If you call a series of equations by the notation M-N, then M and
N must be integers, and N must be greater than or equal to M.}}\else%
 {\count0=#1\count1=#2\advance\count1
 by1\relax\expandafter\@qncall\the\count0,%
  \loop\advance\count0 by1\relax%
    \ifnum\count0<\count1,\expandafter\@qncall\the\count0,%
  \repeat}\fi}

\def\@qneatspace#1#2 {\@qncall#1#2,}
\def\@qncall#1,{\ifunc@lled{#1}{\def\next{#1}\ifx\next\empty\else
  \w@rnwrite{Equation number \noexpand\(>>#1<<) has not been defined yet.}
  >>#1<<\fi}\else\csname @qnnum#1\endcsname\fi}

\let\eqnono=\eqno
\def\eqno(#1){\tag#1}
\def\tag#1$${\eqnono(\displayt@g#1 )$$}

\def\aligntag#1\endaligntag
  $${\gdef\tag##1\\{&(##1 )\cr}\eqalignno{#1\\}$$
  \gdef\tag##1$${\eqnono(\displayt@g##1 )$$}}

\def\eqalignno#1{\displ@y \tabskip\centering
  \halign to\displaywidth{\hfil$\displaystyle{##}$\tabskip\z@skip
    &$\displaystyle{{}##}$\hfil\tabskip\centering
    &\llap{$\displayt@gpar##$}\tabskip\z@skip\crcr
    #1\crcr}}

\def\displayt@gpar(#1){(\displayt@g#1 )}

\def\displayt@g#1 {\rm\ifunc@lled{#1}\global\advance\tagnumber by1
        {\def\next{#1}\ifx\next\empty\else\expandafter
        \xdef\csname @qnnum#1\endcsname{\t@ghead\number\tagnumber}\fi}%
  \writenew@qn{#1}\t@ghead\number\tagnumber\else
        {\edef\next{\t@ghead\number\tagnumber}%
        \expandafter\ifx\csname @qnnum#1\endcsname\next\else
        \w@rnwrite{Equation \noexpand\tag{#1} is a duplicate number.}\fi}%
  \csname @qnnum#1\endcsname\fi}

\def\ifunc@lled#1{\expandafter\ifx\csname @qnnum#1\endcsname\relax}

\let\@qnend=\end\gdef\end{\if@qnfile
\immediate\write16{Equation numbers written on []\jobname.EQN.}\fi\@qnend}

\catcode`@=12
\callall\Equation
\callall\Equations
\callall\Eq
\callall\eq
\callall\Eqs
\callall\eqs


\referencefile

\twelvepoint\doublespace

\title{Hunds Rule Theory for Heavy Fermion
Superconductors}

\author{M. R. Norman}
\affil
Materials Science Division
Argonne National Laboratory
Argonne, IL  60439

\abstract

In this paper, a generalization of standard spin fluctuation theory is
considered which takes into account orbital degeneracy effects which are
critical for describing f electrons.  This theory leads to an
instability for a superconducting pair state which obeys Hunds rules.  Such
a state has L=5, S=1, and J=4.  The degeneracy of this state is broken
by crystalline effects, and realistic calculations for $UPt_3$
find a resultant pair state with $\Gamma_6^-$ symmetry,
consistent with current experimental constraints.

\bigskip

\noindent PACS numbers:  74.20.-z, 74.70.Tx

\endtitlepage

{}From the beginning of theoretical work on heavy fermion superconductors,
it has been realized that there are strong connections between these
metals and superfluid $^3He$.\refto{and2}  This has led many theorists,
including the author, to apply standard spin fluctuation theories which
were developed for $^3He$ to the heavy fermion problem.  So far, the
results have been mixed.  On the plus side, such theories give non
s-wave pairing states, and the evidence in most cases is that the
heavy fermion superconductors are non s-wave.  On the minus side,
the actual group representation these theories predict for $UPt_3$, the
best studied of the heavy fermion superconductors, has so far not
matched what we think the experimental data are telling us.
Available data point to the pair state having $\Gamma_6^-$ ($E_{2u}$)
symmetry.\refto{theory}  This state is an odd parity two dimensional group
representation with line and point nodes, and invariably is
suppressed in the spin fluctuation calculations.\refto{norm}
There are further qualitative problems with these theories.  In Table 1,
a list of the seven known heavy fermion superconductors is shown.  There
are two obvious facts about this table.  Six of the seven superconductors
are uranium alloys.  Second, all of the superconductors either
have two formula units per cell, a point first remarked on by
Anderson,\refto{and1} or have
a magnetic/structural phase transition
at a temperature above $T_c$ so that there are two formula units per cell.
Another interesting point is that the magnetic susceptibilities of the
two heavy fermion superconductors $UPt_3$ and $UPd_2Al_3$ look almost
identical to that of $PrNi_5$, a localized $f^2$ system.  Moreover, the
magnetic susceptibility observed in $URu_2Si_2$ can be easily explained
by an $f^2$ configuration.
The above facts suggest that some on-site interaction is playing
a fundamental role in heavy fermion superconductivity, since such
an interaction could (1) differentiate between Ce and U ions and (2)
depend on having two formula units per cell due to having in phase
or out of phase relations between the order parameters on the two
sites.\refto{and1}  Standard antiferromagnetic spin fluctuation models,
based as they are on having an attractive interaction between near neighbor
sites, do not directly address these points.

The above issues indicate a need to go back and look more closely at
the actual connection between the heavy fermion and $^3He$ cases.
The effective potential of two bare f
electrons on a uranium site looks very similar to the direct interaction
potential for $^3He$.  In particular, the potential is strongly repulsive
at short distances due to the direct Coulomb interaction, is attractive
at intermediate distances (of order 3 a.u.) due to the uranium ion core, and
decays to zero at large distances due to the exponential decay of the f
electron wavefunction.  The ground state of this potential is well known
to have a symmetry of $^3H_4$ (S=1, L=5, J=4) as this state minimizes the
Coulomb repulsion.  This represents a qualitative difference between this
case and that of $^3He$.  In the latter case, one has maximal S also,
but as there is no orbital dependence to the bare interaction, the L state
is fixed by the Landau parameters which are difficult to calculate.  In
the current case, though, the orbital dependence of the interaction
automatically fixes the L state, with the J state being fixed by the
strong spin-orbit interaction.

To understand this problem further, it is useful to review the multiplet
structure for an $f^2$ ion.  The energy terms for the multiplets are best
expressed using Racah parameters (LS scheme).  Every configuration has an
energy $E_0$
which represents the Coulomb repulsion U ($E_0$ is equal to the $L=0$ Coulomb
multipole integral $F^0$ plus a combination of $F^2$, $F^4$, and $F^6$
terms).  The splitting
between singlet and triplet spin states is controlled by the
parameter $E_1$ (a combination of $F^2$, $F^4$, and $F^6$ terms), with
the 3 triplets having a coefficient of 0 and 3 of the 4 singlets have
a coefficient of 2 (the singlet $^1S_0$ is the highest energy state with a
coefficient of 9).  Note the similarity to the paramagnon model for
$^3He$ where the coefficients are the same (0 and 2), but with the important
difference that the splitting in the current case is not determined by
the $F^0$ (charge fluctuation) term as in the single orbital Hubbard model
used for $^3He$ but by the
$L>0$ (shape fluctuation) terms.
Moreover, the degeneracy of the 3 triplets
is lifted by an orbital splitting term, $E_3$, which is another combination
of $L>0$ terms.  The lowest energy state is $^3H$ with an energy $E_0-9E_3$
with the next highest state being $^3F$ with an energy $E_0$.  The $^3F$
energy thus represents the energy of two uncorrelated f electrons and thus
sets the zero of energy (analogous to the $^3He$ problem, where the energy
of the triplet sets the zero of energy).\refto{n1}  Unlike the $^3He$
paramagnon
problem, the
$^3H$ state has an energy lower than the energy zero, i.e., the interaction
is already attractive
at the bare interaction level.  This is only true for a uranium ($f^2$) ion;
for
a cerium ion, the zero of energy is set by the energy of the $f^1$
configuration, and thus the bare interaction is repulsive for all $f^2$
states.
The bare binding energy is equal to the excitation energy
from the $^3H_4$ ground state to the $^3F_2$ state.
This transition has actually been seen
in $UPt_3$ by high energy neutron scattering\refto{osborn} and has an
energy of 0.373 eV.

It is a non-trivial problem to take this bare binding energy and convert
it into an effective binding energy to use between two f electron quasiparticle
states at the Fermi surface.  To do so, collective effects must be included.
First, one takes the bare interaction vertex to be the antisymmetrized Coulomb
interaction (direct minus exchange term).  For the s electron case, one can
easily
show that doing a diagram sum using this bare vertex leads to the standard
paramagnon results.\refto{and3}  The f electron case is more complicated
due to the presence of 4 interaction parameters ($E_0$, $E_1$, $E_2$, $E_3$)
and 14 orbitals.  If one only keeps the $E_0$ term, the diagram series can
be analytically summed.  The result is $E_0/(1-E_0\chi_0)/(1+13E_0\chi_0) +
E_0^2\chi_0/(1-E_0\chi_0)$.
This has some important implications, in that the effective repulsion is
reduced compared to the bare $E_0$ as long as $\chi_0$ is not too close to
being equal to $1/E_0$ (note the difference again to the s electron case, where
the bare repulsion is always reduced in the density channel\refto{and3}).
One can speculate that the divergence for $\chi_0=1/E_0$ represents
a localization instability.
If one keeps only the $E_1$ term, the diagram series can also be summed.
For the triplet states, one obtains $-11E_1^2\chi_0/(1-81E_1^2\chi_0^2)/
(1-4E_1^2\chi_0^2) + 2E_1^2\chi_0/(1-4E_1^2\chi_0^2)$ which has some
similarities to the s electron case.  In
particular, there is an induced attraction for the triplet states.  Note
the divergence for $9E_1\chi_0=1$ which plays the role of a magnetic
instability.  The
analogous series has also been done for the $^1I$ state, where the
induced interaction is $(4E_1+13E_1^2\chi_0-126E_1^3\chi_0^2-162E_1^4\chi_0^3)
/(1-81E_1^2\chi_0^2)/
(1-4E_1^2\chi_0^2) - 2E_1$.  This also has similarities to the s electron
case, and
one finds for singlet states an increased repulsion.
An analytic expression for the general
series has not been obtained due to the complicated coefficients
associated with the $E_2$ and $E_3$ terms.  Instead, one can reduce the vertex
equations to a matrix equation which can be solved on a computer.  This has
been done for the  $^3H$, $^3F$, $^3P$, and $^1I$ states.  Complete orbital
degeneracy has been assumed and no spin-orbit effects have been
included.\refto{n5}

The results are summarized in Figure 1, where the various effective
interactions are plotted versus $\chi_0$ along with the zero of energy
for the $f^1$ case (0) and $f^2$ case ($E_0$).  The values of $E_i$ were gotten
from Goldschmidt\refto{gold} (these values give an $F^0$ term of 1.83 eV,
consistent with spectroscopic data in heavy fermion uranium compounds).  As
one can see, the triplets become increasingly attractive and the singlet
increasingly repulsive as $\chi_0$ increases with a divergence for $(E_0+9E_1)
\chi_0=1$.
$\chi_0$ is difficult to estimate since spin-orbit and anisotropy effects play
a major role.\refto{chi}  For illustrative purposes, we assume a "Stoner"
renormalization of 4 as seen in $^3He$.
For this value of $\chi_0$ (0.137), the $^3H$ energy is -2.3 eV relative to the
$f^2$ zero of energy.  Even for the $f^1$ case, there is still an
(induced) instability for $^3H$, so that pairing is indeed possible for
cerium ions, although less likely.\refto{n4}

One can now estimate the effective pairing matrix element
by realizing that the quasiparticle renormalization in the heavy fermion
case is mostly frequency dependent in nature.\refto{varma}  This would then
act to renormalize the induced interaction discussed above
by a factor of $Z^2$ since each of the four external lines in
the vertex
is renormalized by $Z^{1/2}$ (only Z of the bare f electron is in the
quasiparticle pole).  $Z^{-1}$ is equal to the
mass renormalization factor, known from deHaas-vanAlphen
measurments to be about 16 in $UPt_3$.\refto{dhva1}  This renormalizes the
$^3H$ matrix element of 2.3 eV to
about 100 K.\refto{n3}  This value will be further reduced when
projecting onto pair states at the Fermi energy which have the symmetry of
a particular group representation.  Below, this projection factor is shown
to be about 1/8,
so the final value is 12.5 K.  Since the renormalized (quasiparticle)
Fermi energy, $E_F$, is
about 60 K in $UPt_3$ (specific heat $\gamma$, neutron scattering linewidth),
the pairing coupling constant, $NV_4Z^2$, is about 0.21 (where N is the
renormalized density of states and $V_4$ is the interaction potential in the
$^3H_4$ channel).  With a cut-off of
order $E_F$, this gives a BCS estimate for $T_c$
of 0.6 K.  The agreement with experiment is somewhat fortuitous, of course,
but the point is that the effective coupling constant is at least of the
right order of magnitude.

The actual symmetry of the gap is found by constructing the quasiparticle pair
state $|k,-k>$ using relativistic band structure wavefunctions and projecting
this onto J=4 (this is a specific example of a general procedure advocated
in the past\refto{appel}).  The degeneracy of the J=4 state is broken
due to lattice effects which
should be well described by the momentum dependence of the band structure
wavefunctions (although these wavefunctions fail to describe the
frequency dependence of the quasiparticle states, they give a Fermi surface
shape
in good agreement with experimental data, indicating that their momentum
dependence is reliable).
For hexagonal $UPt_3$, the 18 fold
degeneracy of J=4 in the isotropic case (2J+1 times the number of f sites
in the unit cell, which is two) will
be broken into 3 singlets ($\Gamma_1$, $\Gamma_3$, and
$\Gamma_4$) and 3 doublets (two $\Gamma_5$, one $\Gamma_6$), with each
representation occuring twice (+ (even parity) representations have the two
atoms in phase, and - (odd) have the two atoms out of phase).  In Table 2,
these states are given in terms of pairs of single particle J=5/2 f states.
The group transformation properties of these states are listed in Appel and
Hertel.\refto{appel}

For each k point, there are four
degenerate states available to construct $|k,-k>$ from.\refto{and}  The
singlet (even
parity) combination is
$(|k,Tk> - |PTk,Pk>)/2$ (denoted $d_0$) and the triplet (odd parity)
combinations are $|k,Pk>$ ($-d_x+id_y$), $|PTk,Tk>$ ($d_x+id_y$), and
$(|k,Tk> + |PTk,Pk>)/2$ ($d_z$), where P is the parity operator, and T the
time reversal one.  The odd parity combinations define
a "d" vector which lives in a pseudo-spin space.

The resulting pairing matrix element for this model is then
$$
<k',-k'|H_{eff}|k,-k>_P = (V_4Z^2) A^{*\Gamma\nu j'}_{k'}A^{\Gamma\nu j}_k
\eqno(1)
$$
where P represents the projection with $A^{\Gamma\nu j}_k$ being the
coefficient of the expansion
of $|k,-k>$ which has J=4 with the symmetry of the group representation
$\Gamma$ and basis $\nu$ (for a two-dimensional representation), and j is
the index of the d vector (0 for even, x,y,z for odd).  Since the
matrix element is separable in k and k', it is trivial to write down the
appropriate BCS coupling constant
$$
\lambda = (NV_4Z^2) \sum_j <|A^{\Gamma\nu j}_k|^2>_k
\eqno(2)
$$
where N is the density of states, $< >_k$ is an average over a
narrow energy shell about the Fermi energy, and j runs over 0 for the even
parity case and x,y,z for the odd parity case.

The J=5/2 part of the band structure wavefunctions
can be written as $|k> = \sum a_{\mu i}^{nk}|\mu>_i$ where $\mu$
runs from -5/2
to 5/2, $i$ is the site index (1,2), and n is the band index
(band calculations predict that five f bands contribute to the Fermi surface
of $UPt_3$; such a
surface is in good agreement with deHaas-vanAlphen
data\refto{dhva1}).  Thus, the A coefficients can be written as $\sum
a_{\mu i}^{nk} a_{\nu i'}^{n-k}$ with k denoting either k or PTk
and -k denoting Pk or Tk, with the appropriate linear combinations being
those which match the basis states in Table 2 and have the correct parity
form ($d_0$ for even and $d_x$,$d_y$,$d_z$ for odd).  The average in \Eq{2} was
done by constructing a regular grid of 561 k points in the irreducible wedge
(1/24) of the Brillouin zone and keeping those nk states which are within
1 mRy of the Fermi energy (182 nk points for the current case).
One note is that the coefficients of the $\Gamma_5$ basis
function in
Table 2 are arbitary (subject to normalization).  These are determined
by a variational principle, i.e. one finds the set of coefficients
that gives the maximum coupling constant.

In Table 3, the results of this calculation are given.  The
odd parity states have larger coupling constants since there are
three terms contributing instead of the one term for the even parity case.
This is of interest since the odd parity states
only exist because of the presence of two f atoms in the primitive cell,
which, as mentioned in the introduction, all heavy fermion superconductors
have.
The largest coupling constant
occurs for a state of $\Gamma_6^-$ symmetry.  This state
is an odd parity two dimensional group representation.  It has point
nodes along the c axis and a line of nodes in the $k_z=\pi /c$
zone face.  It is
interesting to note that although only the $d_z$ component of the gap
function vanishes on the $k_z=0$ zone face as expected based on group
theory arguments,\refto{group} all three d vector components vanish on the
$k_z=\pi /c$ zone face, proving a counterexample to the argument in those
papers
that a line node gap function is not possible for odd parity states.
Although the
actual form of the gap function in the current case is extremely complicated
since the $a_{\mu i}^{nk}$ are strong
functions of momentum, this state (1) is from a two dimensional group
representation and can thus explain the unusual phase diagram
seen for $UPt_3$, (2) has the correct nodal structure to explain various
thermodynamic data of $UPt_3$, and (3) is an odd parity state with the largest
possible moment projection onto the basal plane for a two-dimensional group
representation ($M_J=\pm 1$) which is necessary to explain the observed
directional anisotropy of the upper critical field.\refto{choi}  It should be
remarked, though, that the states $\Gamma_1^-$ and $\Gamma_4^-$ have
coupling constants close to that of $\Gamma_6^-$ and the ordering of the
coupling constants will thus be sensitive to the cut-off of the energy shell
used in the averaging in \Eq{2}.  The values
tabulated in Table 3 should be multiplied by the quantity $NV_4Z^2$
to convert to an actual coupling constant, and as discussed above, the
resulting coupling constant for $\Gamma_6^-$ is of the right order to explain
the observed value of $T_c$.  Similar calculations have also been done for
J=2 ($^3F$) and J=0 ($^3P$).  For J=2,
the largest coupling constant also has $\Gamma_6^-$ symmetry (its value modulo
$V_2$ is 0.85 of the J=4 one).  For J=0,
the largest coupling constant has $\Gamma_1^+$ (s-wave) symmetry.
Its value modulo $V_0$ is a factor of four
larger than $J=4$, so it is reassuring to find a repulsive
$V_0$ over a wide range of Figure 1 (in the JJ coupling scheme, $V_2$ and
$V_0$ are repulsive).

In conclusion, an orbital degenerate generalization of the $^3He$ paramagnon
model has been applied to f electrons and yields a superconducting pair state
which satisfies Hunds rules
(L=5, S=1, J=4).  The degeneracy of this state is lifted by crystalline
effects.  Realistic calculations for the case of $UPt_3$ give a pair state
with $\Gamma_6^-$ symmetry which is consistent with experimental data with
a reasonable estimate for $T_c$.  The
theory also explains the preference for heavy fermion superconductors to be
uranium alloys, and also the role that the crystal structure (two formula
units per unit cell) plays in the pairing.

\bigskip

This work was supported by U.S. Department of Energy, Office of Basic Energy
Sciences, under Contract No. W-31-109-ENG-38.  The author would
like to acknowledge the hospitality of the Physics Dept., Uppsala University,
where some of this
work was completed, and to thank Kathyrn Levin for many helpful discussions.

\vfill\eject

\references

\refis{n1} That is, the energy $E_0$ determines the normal state Fermi
energy since two f electrons are occupied per site (the effect of $E_3$
cannot be represented at the single particle level and is assumed to not
enter into this determination).

\refis{n3} One might wonder why such a large interaction would not lead to
a high $T_c$ for more itinerant systems with Z closer to one.  The reason
is that the effective interaction parameters used here are only appropriate for
nearly localized $f^2$ systems.  In palladium, for instance, the effective
interaction is determined by the Stoner I, which leads to a very low estimate
of $T_c$ for triplet superconductivity.

\refis{n4} Actually, the Coulomb repulsion ($E_0$) in cerium is about three
times larger than in uranium, so the tendency towards pairing will be even
further suppressed.

\refis{n5} Details of the vertex equations along with results obtained
with a JJ coupling scheme will be presented in a longer paper.

\refis{chi} $\chi_0$ is actually a momentum dependent matrix in orbital
space.  Anisotropy in the real system will break the rotational invariance
in the vertex equations, leading to an M dependent interaction.  This
complication will be treated in a longer paper.

\refis{norm} M.R. Norman, \prb 48, 6315, 1993; \prb 43, 6121,
1991; \prb 41, 170, 1990.

\refis{and} P.W. Anderson, \prb 30, 4000, 1984; K. Ueda and T. M. Rice,
\prb 31, 7114, 1985.

\refis{and1} P.W. Anderson, \prb 32, 499, 1985.

\refis{and2} P.W. Anderson, \prb 30, 1549, 1984.

\refis{and3} P.W. Anderson and W.F. Brinkman, in The Physics of Liquid
and Solid Helium, Part II, ed. K.H. Bennemann and J.B. Ketterson
(J. Wiley, New York, 1978), p. 177.

\refis{appel} J. Appel and P. Hertel, \prb 35, 155, 1987.

\refis{theory} M.R. Norman, Physica C {\bf 194}, 203 (1992) and references
therein.

\refis{group} G. E. Volovik and L. P. Gor'kov, JETP {\bf 61}, 843 (1985);
E.I. Blount, \prb 32, 2935, 1985.

\refis{choi} C.H. Choi and J.A. Sauls, \prl 66, 484, 1991.

\refis{dhva1} M.R. Norman, R.C. Albers, A.M. Boring, and N.E. Christensen,
Sol. State Comm. {\bf 68}, 245 (1988).

\refis{osborn} R. Osborn, K.A. McEwen, E.A. Goremychkin, A.D. Taylor, Physica
B {\bf 163}, 37 (1990).

\refis{varma} C.M. Varma, \prl 55, 2723, 1985.

\refis{gold} Z.B. Goldschmidt, \pra 27, 740, 1983.

\endreferences

\vfill\eject

\noindent Table 1.  List of known heavy fermion superconductors with the
number of formula units per unit cell.  In parenthesis is the nature of the
low temperature distorted phase in the single formula unit case
(QP - quadrupolar, AF - antiferromagnetic, ? - not fully determined) and
the resulting number of formula units.

\settabs 2 \columns
\vskip12pt
\hrule
\vskip6pt
\+Case&Formula Units\cr
\vskip6pt
\hrule
\vskip6pt
\+$UPt_3$&2\cr
\+$UBe_{13}$&2\cr
\+$U_2PtC_2$&2\cr
\+$URu_2Si_2$&1 (QP/AF - 2)\cr
\+$UPd_2Al_3$&1 (AF - 2)\cr
\+$UNi_2Al_3$&1 (AF - 2)\cr
\+$CeCu_2Si_2$&1 (? - 2 ?)\cr
\vskip6pt
\hrule

\vfill\eject

\noindent Table 2.  Hexagonal basis functions for J=4.  The forms listed in
this table should be (a) antisymmetrized ($|\mu>|\nu>-|\nu>|\mu>$) and
(b) symmetrized (+
representation) or antisymmetrized (- representation) with respect to
site before use.  For $\Gamma_5$, $\alpha$ and $\beta$ are variational
coefficients such that the sum of their squares is equal to one, and
this representation occurs twice ($\alpha,\beta$ and $\beta,-\alpha$).  Note
that both $\Gamma_5$ and $\Gamma_6$ are doublets obtained by replacing
$|\mu>$ by $|-\mu>$.

\settabs 4 \columns
\vskip12pt
\hrule
\vskip6pt
\+Rep&Basis Function\cr
\vskip6pt
\hrule
\vskip6pt
\+$\Gamma_5$&$\alpha |5/2>|3/2> + \beta (0.8018 |5/2>|-1/2> + 0.5976
|3/2>|1/2>)$\cr
\+$\Gamma_3$&$0.7071 |5/2>|1/2> + 0.7071 |-5/2>|-1/2>$\cr
\+$\Gamma_4$&$0.7071 |5/2>|1/2> - 0.7071 |-5/2>|-1/2>$\cr
\+$\Gamma_6$&$0.5345 |5/2>|-3/2> + 0.8452 |3/2>|-1/2>$\cr
\+$\Gamma_1$&$0.2673 (|5/2>|-5/2> +3 |3/2>|-3/2> +2 |1/2>|-1/2>)$\cr
\vskip6pt
\hrule

\vfill\eject

\noindent Table 3.  Coupling constants for J=4 for $UPt_3$.  These are
normalized to the coupling constant for the J=0, $\Gamma_1^+$ (s-wave) state
and should be multiplied by this quantity (0.495, which is the square of
the ratio of the J=5/2 f to the total density of states)
and the quantity $NV_4Z^2$ to convert to real
coupling constants.
Even parity corresponds to + representation
and odd parity to -.

\settabs 3 \columns
\vskip12pt
\hrule
\vskip6pt
\+Rep&even&odd\cr
\vskip6pt
\hrule
\vskip6pt
\+$\Gamma_5$&0.139&0.148\cr
\+$\Gamma_5$&0.059&0.203\cr
\+$\Gamma_3$&0.048&0.129\cr
\+$\Gamma_4$&0.027&0.242\cr
\+$\Gamma_6$&0.036&0.253\cr
\+$\Gamma_1$&0.153&0.229\cr
\vskip6pt
\hrule

\vfill\eject

\figurecaptions

\figis{1} Effective interaction (eV) for $^3H$, $^3F$, $^3P$, and $^1I$
versus $\chi_0$ for parameters appropriate to a U
ion\refto{gold} ($E_0$ = 1225 meV,
$E_1$ = 470.3 meV, $E_2$ = 1.923 meV, $E_3$ = 43.28 meV).  The zeros of
energy for the $f^1$ and $f^2$ cases are marked by the dashed lines.

\endfigurecaptions

\vfill\eject

\endit

XXX

userdict/LW{save statusdict/product get(LaserWriter)anchorsearch
exch pop{dup length 0 eq{pop 1}{( Plus)eq{2}{3}ifelse}ifelse}{0}ifelse
 exch restore}bind put
userdict/downloadOK known not{userdict/downloadOK{systemdict dup/eexec
 known exch/cexec known and LW dup 1 ne exch 2 ne and and vmstatus exch
 sub exch pop 120000 gt and}bind put}if
userdict/type42known known not{userdict/type42known
 systemdict/resourcestatus known{42/FontType resourcestatus{pop pop true
}{false}ifelse }{false}ifelse put}if
type42known not downloadOK and {userdict begin /*charpath /charpath load
 def/charpathflag false def/charpath{userdict/charpathflag true put
 userdict/*charpath get exec userdict/charpathflag false put}bind def
 end}if
userdict/checkload known not{userdict/checkload{{pop exec} {save 3 dict
 begin/mystring 6050 string def
exch/endstring exch def{currentfile mystring readline not{stop}if
 endstring eq{exit}if}loop end restore pop}ifelse}bind put}if
userdict/LW+{LW 2 eq}bind put
userdict/ok known not{userdict/ok{systemdict/statusdict known dup{LW 0
 gt and}if}bind put}if
systemdict/currentpacking known{currentpacking true setpacking}if
/md 270 dict def md begin
/av 71 def
/T true def/F false def/mtx matrix def/s75 75 string def/sa8 8 string
 def/sb8 8 string def
/sc8 8 string def/sd8 8 string def/s1 ( ) def/pxs 1 def/pys 1 def
/ns false def
1 0 mtx defaultmatrix dtransform exch atan/pa exch def/nlw .24 def/ppr
 [-32 -29.52 762 582.48] def
/pgr [0 0 0 0] def
/pgs 1 def/por true def/xb 500 array def/so true def/tso true
 def/fillflag false def/pnm 1 def/fmv true def
/sfl false def/ma 0 def/invertflag false def/dbinvertflag false
 def/xflip false def/yflip false def/noflips true def/scaleby96 false
 def/fNote true def/fBitStretch true def
/4colors false def/fg (Rvd\001\001\000\000\177) def
/bdf{bind def}bind def
/xdf{exch def}bdf
/xl{neg exch neg translate}bdf
/fp{pnsh 0 ne pnsv 0 ne and}bdf
/nop{}bdf/lnop[/nop load]cvx bdf
/vrb[
{fp{fg 6 get 0 ne{gsave stroke grestore}{gsave 1 setlinewidth pnsh pnsv
 scale stroke grestore}ifelse}if newpath}bind
/eofill load
dup
/newpath load
2 index
dup
{clip newpath}bind
{}bind
dup
2 copy
]def
/sgd systemdict/setpagedevice known{{2 dict begin/PreRenderingEnhance
 exch def/Policies 1 dict dup/PreRenderingEnhance 1 put def currentdict
 end setpagedevice}}{{pop}}ifelse bdf

/svsc systemdict/currentcolorscreen known{{currentcolorscreen/dkspf
 xdf/dkrot xdf/dkfreq xdf/dyspf xdf/dyrot xdf/dyfreq xdf/dmspf xdf/dmrot
 xdf/dmfreq xdf
/dcspf xdf/dcrot xdf/dcfreq xdf}}{{currentscreen/spf xdf/rot xdf/freq
 xdf}}ifelse bdf
/doop{vrb exch get exec}bdf
/psu{/udf xdf/tso xdf /fNote xdf/fBitStretch xdf/scaleby96 xdf/yflip
 xdf/xflip xdf
/invertflag xdf/dbinvertflag invertflag statusdict begin version cvr
 47.0 ge product (LaserWriter) eq not and end invertflag and {not}if def
xflip yflip or{/noflips false def}if
/pgs xdf 2 index .72 mul exch div/pys xdf div .72 mul/pxs xdf ppr astore
 pop pgr astore pop/por xdf sn and/so xdf}bdf
/tab{userdict /11x17 known{userdict begin /11x17 load exec end}
{statusdict /setpage known{statusdict begin 792 1224 1 setpage end}
{statusdict /setpageparams known{statusdict begin 792 1224 0 1
 setpageparams end}if}ifelse}ifelse}bdf
/a3Size{userdict /a3 known{userdict begin /a3 load exec end}{statusdict
 /setpageparams known{statusdict begin 842 1191 0 1 setpageparams end}if
}ifelse}bdf
/txpose{fNote{smalls}{bigs}ifelse pgs get exec pxs pys scale ppr aload
 pop por{noflips{pop exch neg exch translate pop 1 -1 scale}if
xflip yflip and{pop exch neg exch translate 180 rotate 1 -1 scale ppr 3
 get ppr 1 get neg sub neg ppr 2 get ppr 0 get neg sub neg translate}if
xflip yflip not and{pop exch neg exch translate pop 180 rotate ppr 3 get
 ppr 1 get neg sub neg 0 translate}if yflip xflip not and{ppr 1 get neg
 ppr 0 get neg translate}if}
{noflips{translate pop pop 270 rotate 1 -1 scale}if xflip yflip and
{translate pop pop 90 rotate 1 -1 scale ppr 3 get ppr 1 get neg sub neg
 ppr 2 get ppr 0 get neg sub neg translate}if
xflip yflip not and{translate pop pop 90 rotate ppr 3 get ppr 1 get neg
 sub neg 0 translate}if yflip xflip not and{translate pop pop 270 rotate
 ppr 2 get ppr 0 get neg sub neg 0 exch translate}if}ifelse
statusdict begin/waittimeout where{pop waittimeout 300 lt
{statusdict/waittimeout 300 put}if}if end
scaleby96{ppr aload pop 4 -1 roll add 2 div 3 1 roll add 2 div 2 copy
 translate .96 dup scale neg exch neg exch translate}if}bdf
/fr{4 copy pgr aload pop 3 -1 roll add 3 1 roll exch add 6 2 roll 3 -1
 roll
sub 3 1 roll exch sub 3 -1 roll exch div 3 1 roll div exch scale pop pop
 xl}bdf
/obl{{0.212557 mul}{pop 0}ifelse}bdf
/sfd{ps fg 5 -1 roll get mul 100 div 0 ps 5 -1 roll obl ps neg 0 0 6a
 astore makefont setfont}bdf
/fnt{findfont sfd}bdf
/bt{sa 3 1 roll 3 index and put}bdf
/sa(\000\000\000\000\000\000\000\000\000\000)def
/fs{0 1 bt 1 2 bt 2 4 bt 3 8 bt 4 16 bt 5 32 bt 6 64 bt 7 128 bt sa exch
 8 exch put}bdf
/mx1 matrix def
/mx2 matrix def
/mx3 matrix def
/bu{currentpoint 4colors{currentcmykcolor}{currentrgbcolor}ifelse
 currentlinewidth currentlinecap currentlinejoin
currentdash exch aload length fg 5 sfl{1}{0}ifelse put pnsv pnsh
2t aload pop 3a aload pop mx2 aload pop mx1 aload pop mtx currentmatrix
 aload pop
mx3 aload pop ps pm restore/ps xdf mx3 astore pop}bdf
/bn{/pm save def mx3 setmatrix newpath 0 0 moveto ct dup 39 get 0 exch
 getinterval cvx exec mtx astore setmatrix mx1 astore pop mx2 astore pop
 3a
astore pop 2t astore pop/pnsh xdf/pnsv xdf gw
/sfl fg 5 get 0 ne def array astore exch setdash setlinejoin setlinecap
setlinewidth 4colors{mysetcmykcolor}{setrgbcolor}ifelse moveto}bdf
/fc{save vmstatus exch sub 50000 lt
{(
/tc{32768 div add 3 1 roll 32768 div add 2t astore pop}bdf
/3a [0 0 0] def
/2t 2 array def
/tp{3a astore pop}bdf
/tt{mx2 currentmatrix pop currentpoint 2 copy 2t aload pop qa 2 copy
 translate 3a aload pop exch dup 0 eq
{pop}{1 eq{-1 1}{1 -1}ifelse scale}ifelse rotate pop neg exch neg exch
 translate moveto}bdf
/te{mx2 setmatrix}bdf
/th{3 -1 roll div 3 1 roll exch div 2 copy mx1 scale pop scale/sfl true
 def}bdf
/tu{1 1 mx1 itransform scale/sfl false def}bdf
/ts{1 1 mx1 transform scale/sfl true def}bdf
/fz{/ps xdf}bdf
/dv{dup 0 ne{div}{pop}ifelse}bdf
/pop4{pop pop pop pop}bdf
/it{sfl{mx1 itransform}if}bdf
/gm{exch it moveto}bdf/rm{it rmoveto}bdf
/lm{currentpoint sfl{mx1 transform}if exch pop sub 0 exch it rmoveto}bdf
/fm{statusdict/manualfeed known}bdf
/se{statusdict exch/manualfeed exch put}bdf
/mf{dup/ma exch def 0 gt{fm se/t1 5 st ok ma 1 gt and{/t2 0 st/t3 0 st
statusdict/manualfeedtimeout 3600 put
}if}if}bdf
/jn{/statusdict where exch pop{statusdict exch /jobname exch put}if}bdf
/pen{pnm mul/pnsh xdf pnm mul/pnsv xdf pnsh setlinewidth}bdf
/min{2 copy gt{exch}if pop}bdf
/max{2 copy lt{exch}if pop}bdf
/dh{fg 6 1 put array astore dup {1 pxs div mul exch}forall astore exch
 pop exch pop exch setdash}bdf
/ih[currentdash]def
/rh{fg 6 0 put ih aload pop setdash}bdf
/dl{gsave nlw pys div setlinewidth 0 setgray}bdf
/dlin{exch currentpoint currentlinewidth 2 div dup
translate newpath moveto lineto currentpoint stroke grestore moveto}bdf
/lin{fg 6 get 0 ne{exch lineto currentpoint 0 doop moveto}
{exch currentpoint/pnlv xdf/pnlh xdf gsave newpath/@1 xdf/@2 xdf fp{pnlh
 @2 lt{pnlv @1 ge
{pnlh pnlv moveto @2 @1 lineto pnsh 0 rlineto
0 pnsv rlineto pnlh pnsh add pnlv pnsv add lineto pnsh neg 0 rlineto}
{pnlh pnlv moveto pnsh 0 rlineto @2 pnsh add @1 lineto 0 pnsv rlineto
pnsh neg 0 rlineto pnlh pnlv pnsv add lineto}ifelse}{pnlv @1 gt
{@2 @1 moveto pnsh 0 rlineto pnlh pnsh add pnlv lineto 0 pnsv rlineto
pnsh neg 0 rlineto @2 @1 pnsv add lineto}{pnlh pnlv moveto pnsh 0
 rlineto
0 pnsv rlineto @2 pnsh add @1 pnsv add lineto pnsh neg 0 rlineto
0 pnsv neg rlineto}ifelse}ifelse
closepath fill}if @2 @1 grestore moveto}ifelse}bdf
/gw{/pnm fg 3 get fg 4 get div def}bdf
/lw{fg exch 4 exch put fg exch 3 exch put gw pnsv pnsh pen}bdf
/barc{/@1 xdf/@2 xdf/@3 xdf/@4 xdf/@5 xdf
/@6 xdf/@7 xdf/@8 xdf gsave
@5 @7 add 2 div @6 @8 add 2 div translate newpath 0 0 moveto
@5 @7 sub @6 @8 sub mtx currentmatrix pop scale @1{newpath}if
0 0 0.5 @4 @3 arc @4 @3 sub abs 360 ge{closepath}if
mtx setmatrix @2 doop grestore}bdf
/ar{dup 0 eq barc}bdf
/ov{0 exch 360 exch true barc}bdf
/rc{dup/@t xdf 0 eq{4 copy 3 -1 roll eq 3 1 roll eq and{pnsv 2 div sub
 exch pnsh 2 div sub exch 4 2 roll pnsv 2 div add exch pnsh 2 div add
 exch
/@t 1 def}if}if currentpoint 6 2 roll newpath 4 copy 4 2 roll exch
 moveto 6 -1 roll lineto lineto lineto closepath @t doop moveto}bdf
/mup{dup pnsh 2 div le exch pnsv 2 div le or}bdf
/rr{/@1 xdf 2. div/@2 xdf 2. div/@3 xdf
/@4 xdf/@5 xdf/@6 xdf/@7 xdf
@7 @5 eq @6 @4 eq @2 mup or or{@7 @6 @5 @4 @1 rc}
{@4 @6 sub 2. div dup @2 lt{/@2 xdf}{pop}ifelse
@5 @7 sub 2. div dup @2 lt{/@2 xdf}{pop}ifelse
@1 0 eq{/@2 @2 pnsh 2 div 2 copy gt{sub def}{0 pop4}ifelse}if
currentpoint newpath
@4 @6 add 2. div @7 moveto
@4 @7 @4 @5 @2 arcto pop4
@4 @5 @6 @5 @2 arcto pop4
@6 @5 @6 @7 @2 arcto pop4
@6 @7 @4 @7 @2 arcto pop4
closepath @1 doop moveto}ifelse}bdf
/pr{gsave newpath/pl{exch moveto/pl{exch lineto}def}def}bdf
/pl{exch lineto}bdf
/ep{dup 0 eq{{moveto}{exch lin}{}{(
pop grestore}{doop grestore}ifelse currentpoint newpath moveto}bdf
/gr{64. div setgray}bdf
/savescreen{ns not{/ns true def systemdict/currentcolorscreen known
{currentcolorscreen/pkspf xdf/pkrot xdf/pkfreq xdf/pyspf xdf/pyrot
 xdf/pyfreq xdf/pmspf xdf/pmrot xdf/pmfreq xdf
/pcspf xdf/pcrot xdf/pcfreq xdf}{currentscreen/sspf xdf/srot xdf/sfreq
 xdf}ifelse}if}bdf
/restorescreen{/ns false def systemdict/setcolorscreen known{pcfreq
 pcrot/pcspf load pmfreq pmrot/pmspf load pyfreq pyrot/pyspf load
pkfreq pkrot/pkspf load setcolorscreen}{sfreq srot/sspf load setscreen
}ifelse}bdf
/pat{savescreen sa8
copy pop 9.375 pa por not{90 add}if{1 add 4 mul cvi sa8 exch get exch 1
 add 4 mul cvi 7 sub bitshift 1 and}setscreen exch not{gr}{pop}ifelse
}bdf
/sg{restorescreen gr}bdf
/cpat{savescreen 10 2 roll 7 -1 roll sa8 copy pop 9.375 pa por not{90
 add}if{1 add 4 mul cvi sa8 exch get exch 1 add 4 mul cvi 7 sub bitshift
 1 and}8 -1 roll sb8 copy pop 9.375 pa por not{90 add}if{1 add 4 mul cvi
 sb8
exch get exch 1 add 4 mul cvi 7 sub bitshift 1 and}9 -1 roll sc8 copy
 pop 9.375 pa por not{90 add}if{1 add 4 mul cvi sc8 exch get exch 1 add
 4 mul cvi 7 sub bitshift 1 and}10 -1 roll sd8 copy pop 9.375 pa por not
{90 add}if{1 add 4 mul cvi sd8
exch get exch 1 add 4 mul cvi 7 sub bitshift 1 and}psuedo1 dsc 4{4 -1
 roll 1 exch 64 div sub}repeat mysetcmykcolor pop pop}bdf
systemdict/setcolorscreen known{/psuedo1 lnop bdf/dsc/setcolorscreen
 load def}{/psuedo1{16{pop}repeat sa8 copy pop 9.375 pa por not{90 add
}if{1 add 4 mul cvi sa8 exch get exch 1 add 4 mul cvi 7 sub bitshift 1
 and}}bdf
/bwsc{setscreen dup gr 0 exch 0 exch 64 exch 64 exch 64 exch
}bdf/dsc/bwsc load def
}ifelse
systemdict/setcmykcolor known{/mysetcmykcolor /setcmykcolor load def}
{/mysetcmykcolor{1 sub 4 1 roll 3{3 index add neg dup 0 lt{pop 0}if 3 1
 roll}repeat setrgbcolor pop}bdf}ifelse
/dc{transform round .5 sub exch round .5 sub exch itransform}bdf
/sn{userdict/smooth4 known}bdf
/x8{3 bitshift}bdf
/x4{2 bitshift}bdf
/d4{-2 bitshift}bdf
/d8{-3 bitshift}bdf
/rb{15 add -4 bitshift 1 bitshift}bdf
/db{/@7 save def/@1 xdf/@2 xdf/@3 xdf/@4 xdf/@5 xdf/@6 @5 @3 4 add mul
 def
dc translate scale/xdbit 1 1 idtransform abs/ydbit exch def abs def{0 0
 1 ydbit add 1 10 rc clip}if
@1 0 eq @1 4 eq or{currentrgbcolor 1 setgray ydbit 0 1 ydbit add 1 2 rc
 setrgbcolor}if
@1 3 eq @1 7 eq or{1 setgray}{currentrgbcolor 2 index eq exch 2 index eq
 and exch pop{0 setgray}if}ifelse/@9 @1 0 eq @1 1 eq @1 3 eq or or
 dbinvertflag xor def/@13 @6 def
@2 fBitStretch or{/@10 @4 x4 def/@11 @3 x4 def/@12 @10 rb def/@13 @12
 @11 mul def/@15 1 1 dtransform abs/calcY 1 index def round cvi/@14 exch
 def
abs/calcX 1 index def round cvi scaleby96 not{1 add}if def/@16 @15 rb
 def/@17 @16 @14 mul def}if
sn @13 60000 lt and @2 fBitStretch or and{mtx currentmatrix dup 1 get
 exch 2 get 0. eq exch 0. eq and @17 60000 lt and fBitStretch and{@16 3
 bitshift @14 @9 [calcX 0 0 calcY 0 0]{@17 string @13 string
currentfile @6 string readhexstring pop 1 index @4 @3 @5 @12 @2 smooth4
@10 @11 @12 dup string 5 index @15 @14 @16 dup string stretch}imagemask}
{@12 x8 @11 @9 [@10 0 0 @11 0 0]{@13 string
currentfile @6 string readhexstring pop 1 index @4 @3 @5 @12 @2 smooth4
}imagemask}ifelse}{@5 3 bitshift @3 4 add @9 [@4 0 0 @3 0 2]{currentfile
 @6 string readhexstring pop}imagemask}ifelse
@7 restore}bdf
systemdict/setcmykcolor known{/psuedo lnop bdf/di/colorimage load def}
{/routines[{.3 mul add 1}bind{.59 mul add 2}bind{.11 mul add round cvi
 str exch i exch put/i i 1 add def 0 0}bind]def
/psuedo{/i 0 def 0 exch 0 exch{exch routines exch get exec}forall pop
 pop str}bdf/bwi{pop pop image}bdf/di/bwi load def}ifelse
/cdb{/@7 save def/@1 xdf/@2 xdf/@3 xdf/@4 xdf/@5 xdf
systemdict/setcmykcolor known not{dc}if translate scale /@6 xdf
/@18 @5 dup 60000 ge{pop 60000}if string def @6 not{/str @18 0 @18
 length 3 idiv getinterval def}if @4 @3 8 [@4 0 0 @3 0 0]@6{{currentfile
 @18 readhexstring pop}image}{{currentfile @18 readhexstring pop psuedo
}false 3 di}ifelse @7 restore}bdf
/wd 16 dict def
/mfont 14 dict def
/mdf{mfont wcheck not{/mfont 14 dict def}if mfont begin xdf end}bdf
/cf{{1 index/FID ne{def}{pop pop}ifelse}forall}bdf/rf{/@1 exch def/@2
 exch def
FontDirectory @2 known{cleartomark pop}{findfont dup begin dup length @1
 add dict begin
cf{/Encoding macvec def}{Encoding dup length array copy/Encoding exch
 def
counttomark 2 idiv{Encoding 3 1 roll put}repeat}ifelse
pop
exec currentdict end end @2 exch definefont pop}ifelse}bdf
/bmbc{exch begin wd begin
/cr xdf
save
CharTable cr 6 mul 6 getinterval{}forall
/bitheight xdf/bitwidth xdf
.96 div/width xdf
Gkernmax add/XOffset xdf Gdescent add/YOffset xdf/rowbytes xdf
rowbytes 255 eq{0 0 0 0 0 0 setcachedevice}
{Gnormsize dup scale
width 0 XOffset YOffset bitwidth XOffset add bitheight YOffset add
setcachedevice
rowbytes 0 ne{
XOffset YOffset translate newpath 0 0 moveto
bitwidth bitheight scale
sn{
/xSmt bitwidth x4 def
/ySmt bitheight x4 def
/rSmt xSmt rb def
rSmt x8 ySmt true
[xSmt 0 0 ySmt neg 0 ySmt]
{rSmt ySmt mul string CharData cr get
1 index bitwidth bitheight rowbytes rSmt tso smooth4}
}{rowbytes 3 bitshift bitheight 4 add true
[bitwidth 0 0 bitheight neg 0 bitheight 2 add]
{CharData cr get}
}ifelse
imagemask
}if
}ifelse
restore
end end
}bdf
/bb{.96 exch div/Gnormsize mdf 2 index
/Gkernmax mdf 1 index/Gdescent mdf
3 index div 4 1 roll
2 index div 1. 5 2 roll
exch div 4 1 roll
4 array astore/FontBBox mdf
}bdf
/cdf{mfont/CharData get 3 1 roll put}bdf
/bf{
mfont begin
/FontType 3 def
/FontMatrix [1 0 0 1 0 0] def
/Encoding macvec def
/MFontType 0 def
/BuildChar/bmbc load def
end
mfont definefont pop
}bdf
/wi LW 1 eq{{gsave 0 0 0 0 0 0 0 0 moveto lineto lineto lineto closepath
 clip stringwidth grestore}bind}{/stringwidth load}ifelse def
/aps{0 get 124 eq}bdf
/xc{s75 cvs dup}bdf
/xp{put cvn}bdf
/scs{xc 3 67 put dup 0 95 xp}bdf
/sos{xc 3 79 xp}bdf
/sbs{xc 1 66 xp}bdf
/sis{xc 2 73 xp}bdf
/sob{xc 2 79 xp}bdf
/sss{xc 4 83 xp}bdf
/dd{exch 1 index add 3 1 roll add exch}bdf
/smc{moveto dup show}bdf
/ndf2{udf{dup /FontType get 0 eq{/FDepVector get{dup /FontType get 0 eq
{ndf2}{dup /df2 known{begin df2 0 null put end
}{pop}ifelse}ifelse}forall}{/df2 known{dup begin df2 0 null put end}if
}ifelse}{pop}ifelse}bdf
/kwn{FontDirectory 1 index known{findfont dup ndf2 exch pop}}bdf
/gl{1 currentgray sub setgray}bdf
/newmm{dup /FontType get 0 eq{dup maxlength dict begin{1 index/FID ne 2
 index/UniqueID ne and{def}{pop pop}ifelse}forall currentdict end
dup /FDepVector 2 copy get[exch 6 index exch 6 index exch{newmm 3 1 roll
}forall pop pop] put dup
}{/mfont 10 dict def mfont begin/FontMatrix [1 0 0 1 0 0] def
/FontType 3 def/Encoding macvec def/df 1 index def/df2 1 array
 def/FontBBox [0 0 1 1] def/StyleCode 2 index def
/mbc{bcarray StyleCode get}def/BuildChar{exch begin	wd begin/cr exch
 def/cs s1 dup 0 cr put def df /MFontType known not{
df2 0 get null eq{df dup length 2 add dict begin{1 index/FID ne 2
 index/UniqueID ne and{def}{pop pop}ifelse}forall
/StrokeWidth 1 0 FontMatrix idtransform pop dup nlw mul pys div ps div
 exch 0.012 mul 2 copy le{exch}if pop def/PaintType 2 def currentdict
 end
/q exch definefont df2 exch 0 exch put}if}if mbc exec end end}def end
 mfont}ifelse
3 index exch definefont exch pop}bdf
/mb{dup sbs kwn{0 2 index findfont newmm exch pop exch pop exch pop
}ifelse sfd}bdf
/mo{dup sos kwn{2 2 index findfont newmm exch pop exch pop exch pop
}ifelse sfd}bdf
/ms{dup sss kwn{4 2 index findfont newmm exch pop exch pop exch pop
}ifelse sfd}bdf
/ou{dup sos kwn{mfont/df2 known{mfont begin df2 0 null put end}if 3 2
 index findfont newmm exch pop exch pop exch pop}ifelse sfd}bdf
/su{dup sss kwn{mfont/df2 known{mfont begin df2 0 null put end}if 5 2
 index findfont newmm exch pop exch pop exch pop}ifelse sfd}bdf
/ao{/fmv true def ou}bdf/as{/fmv true def su}bdf
/vo{/fmv false def ou}bdf/vs{/fmv false def su}bdf
/c{currentrgbcolor dup 4 1 roll eq 3 1 roll eq and/gray xdf}bdf
/bcarray[{/da .03 def df setfont gsave cs wi 1 index 0 ne{exch da add
 exch}if grestore setcharwidth
cs 0 0 smc da 0 smc da da smc 0 da moveto show}bind dup{/da 1 ps div def
 df setfont gsave cs wi 1 index 0 ne{exch da add exch}if grestore
 setcharwidth
cs 0 0 smc da 0 smc da da smc 0 da smc c gray{gl}{1 setgray}ifelse da 2.
 div dup moveto show}bind
{df setfont gsave cs wi grestore setcharwidth c gray{gl}{currentrgbcolor
 1 setgray}ifelse cs 0 0 smc df2 0 get setfont
gray{gl}{4 1 roll setrgbcolor}ifelse 0 0 moveto show}bind
{/da 1 ps div def/ds .05 def/da2 da 2. div def df setfont gsave cs wi 1
 index 0 ne{exch ds add da2 add exch}if grestore setcharwidth
cs ds da2 add .01 add 0 smc 0 ds da2 sub translate 0 0 smc da 0 smc da
 da smc 0 da smc c gray{gl}{1 setgray}ifelse da 2. div dup moveto show
}bind
{/da .05 def df setfont gsave cs wi 1 index 0 ne{exch da add exch}if
 grestore setcharwidth c cs da .01 add 0 smc 0 da translate
gray{gl}{currentrgbcolor 1 setgray 4 -1 roll}ifelse 0 0 smc gray{gl}{4 1
 roll setrgbcolor}ifelse df2 0 get setfont 0 0 moveto show}bind]def
/st{1000 mul usertime add dup 2147483647 gt{2147483647 sub}if def}bdf
/the{usertime sub dup 0 lt exch -2147483648 gt and}bdf
/6a 6 array def
/2a 2 array def
/3q 3 array def
/qs{3 -1 roll sub exch 3 -1 roll sub exch}bdf
/qa{3 -1 roll add exch 3 -1 roll add exch}bdf
/qm{3 -1 roll 1 index mul 3 1 roll mul}bdf
/qn{6a exch get mul}bdf
/qA .166667 def/qB .833333 def/qC .5 def
/qx{6a astore pop
qA 0 qn qB 2 qn add   qA 1 qn qB 3 qn add
qB 2 qn qA 4 qn add   qB 3 qn qA 5 qn add
qC 2 qn qC 4 qn add   qC 3 qn qC 5 qn add}bdf
/qp{6 copy 12 -2 roll pop pop}bdf
/qc{exch qp qx curveto}bdf
/qi{{exch 4 copy 2a astore aload pop qa .5 qm newpath moveto}{exch 2
 copy 6 -2 roll 2 qm qs 4 2 roll}ifelse}bdf
/qq{{qc 2a aload pop qx curveto}{exch 4 copy qs qa qx curveto}ifelse}bdf
/pt{currentpoint newpath moveto}bdf
/qf{/fillflag true def}bdf
/ec{dup 4 and 0 ne{closepath}if 1 and 0 ne{0 doop}if grestore
 currentpoint newpath moveto/fillflag false def}bdf
/eu{currentpoint fp{0 ep}{grestore newpath}ifelse moveto/fillflag false
 def}bdf
/bp{currentpoint newpath 2 copy moveto}bdf
/ef{gsave fillflag{gsave eofill grestore}if}bdf
/sm{0 exch{@1 eq{1 add}if}forall}bdf
/lshow{4 1 roll exch/@1 exch def{1 index wi pop sub 1 index sm dv 0 @1 4
 -1 roll widthshow}{1 index wi pop sub
1 index dup sm 10 mul exch length 1 sub add dv dup 10. mul 0 @1 4 -1
 roll 0 6 -1 roll awidthshow}ifelse}bdf
/setTxMode{sa 9 2 index put exch not{3 eq{1}{0}ifelse setgray}{pop
}ifelse}bdf
/SwToSym{{}mark false/Symbol/|______Symbol 0 rf 0 sa 6 get 0 ne{pop 1}
{sa 7 get 0 eq{pop 2}if}ifelse
sa 1 get 0 ne/|______Symbol
sa 4 get 0 ne{vs}{sa 3 get 0 ne{vo}{fnt}ifelse}ifelse}bdf
/mc{0 3 1 roll transform neg exch pop}bdf
/ul{dup 0 ne sa 2 get 0 ne and{gsave 0 0
/UnderlinePosition kif{mc}{ps -10 div}ifelse/UnderlineThickness kif{mc}
{ps 15 div}ifelse
abs setlinewidth neg rmoveto
sa 4 get 0 ne{gsave currentlinewidth 2. div dup rmoveto currentpoint
 newpath moveto
2 copy rlineto stroke grestore}if
sa 3 get sa 4 get or 0 ne{gsave currentrgbcolor dup 4 1 roll eq 3 1 roll
 eq and{gl}{1 setgray}ifelse 2 copy rlineto stroke grestore rlineto
 strokepath nlw pys div setlinewidth}{rlineto}ifelse
stroke grestore}{pop}ifelse}bdf
/sgt{2 copy known{get true}{pop pop false}ifelse}bdf
/kif{currentfont dup/FontMatrix get exch/FontInfo sgt{true}
{currentfont/df sgt
{dup/FontInfo sgt{3 1 roll/FontMatrix get mtx concatmatrix exch true}
{pop pop pop false}
ifelse}{pop pop false}ifelse}ifelse{3 -1 roll sgt{exch true}{pop false
}ifelse}{false}ifelse}bdf
/blank/Times-Roman findfont/CharStrings get/space get def
/macvec 256 array def
/NUL/SOH/STX/ETX/EOT/ENQ/ACK/BEL/BS/HT/LF/VT/FF/CR/SO/SI
/DLE/DC1/DC2/DC3/DC4/NAK/SYN/ETB/CAN/EM/SUB/ESC/FS/GS/RS/US
macvec 0 32 getinterval astore pop
macvec 32/Times-Roman findfont/Encoding get
32 96 getinterval putinterval macvec dup 39/quotesingle put 96/grave put
/Adieresis/Aring/Ccedilla/Eacute/Ntilde/Odieresis/Udieresis/aacute
/agrave/acircumflex/adieresis/atilde/aring/ccedilla/eacute/egrave
/ecircumflex/edieresis/iacute/igrave/icircumflex/idieresis/ntilde/oacute
/ograve/ocircumflex/odieresis/otilde/uacute/ugrave/ucircumflex/udieresis
/dagger/degree/cent/sterling/section/bullet/paragraph/germandbls
/registered/copyright/trademark/acute/dieresis/notequal/AE/Oslash
/infinity/plusminus/lessequal/greaterequal/yen/mu/partialdiff/summation
/product/pi/integral/ordfeminine/ordmasculine/Omega/ae/oslash
/questiondown/exclamdown/logicalnot/radical/florin/approxequal/Delta/guillemotleft
/guillemotright/ellipsis/blank/Agrave/Atilde/Otilde/OE/oe
/endash/emdash/quotedblleft/quotedblright/quoteleft/quoteright/divide/lozenge
/ydieresis/Ydieresis/fraction/currency/guilsinglleft/guilsinglright/fi/fl
/daggerdbl/periodcentered/quotesinglbase/quotedblbase/perthousand/Acircumflex/Ecircumflex/Aacute
/Edieresis/Egrave/Iacute/Icircumflex/Idieresis/Igrave/Oacute/Ocircumflex
/apple/Ograve/Uacute/Ucircumflex/Ugrave/dotlessi/circumflex/tilde
/macron/breve/dotaccent/ring/cedilla/hungarumlaut/ogonek/caron
macvec 128 128 getinterval astore pop
{}mark true/Courier/|______Courier 0 rf
{/Metrics 21 dict begin/zero 600 def/one 600 def/two 600 def/three 600
 def/four 600 def/five 600 def/six 600 def/seven 600 def/eight 600 def
/nine 600 def/comma 600 def/period 600 def/dollar 600 def/numbersign 600
 def/percent 600 def/plus 600 def/hyphen 600 def/E 600 def/parenleft 600
 def/parenright 600 def/space 600 def
currentdict end def currentdict/UniqueID known{/UniqueID 16#800000 def
}if/FontBBox FontBBox 4 array astore def}mark
 true/Helvetica/|______Seattle 1 rf
/oldsettransfer/settransfer load def
/concatprocs{/proc2 exch cvlit def/proc1 exch cvlit def/newproc proc1
 length proc2 length add array def
newproc 0 proc1 putinterval newproc proc1 length proc2 putinterval
 newproc cvx}def
/settransfer{currenttransfer concatprocs oldsettransfer}def
/PaintBlack{{1 exch sub}settransfer gsave newpath clippath 1 setgray
 fill grestore}def
/od{(Rvd\001\001\000\000\177) fg copy pop txpose
1 0 mtx defaultmatrix dtransform exch atan/pa exch def
newpath clippath mark
{transform{itransform moveto}}{transform{itransform lineto}}
{6 -2 roll transform 6 -2 roll transform 6 -2 roll transform
{itransform 6 2 roll itransform 6 2 roll itransform 6 2 roll curveto}}
{{closepath}}pathforall newpath counttomark array astore/gc xdf pop ct
 39 0 put
10 fz 0 fs 2 F/|______Courier fnt invertflag{PaintBlack}if
statusdict/processcolors known{statusdict begin processcolors end 4 eq
{/4colors true def}if}if}bdf
/cd{}bdf
/op{/sfl false def systemdict/currentcolorscreen known{dcfreq
 dcrot/dcspf load dmfreq dmrot/dmspf load dyfreq dyrot/dyspf load
dkfreq dkrot/dkspf load setcolorscreen}{freq rot/spf load setscreen
}ifelse savescreen
/ns false def/pm save def}bdf
/cp{not{userdict/#copies 0 put}if ma 0 gt{{t1 the{exit}if}loop}if
{/copypage load exec}{/showpage load exec}ifelse pm restore}bdf
/px{0 3 1 roll tp tt}bdf
/psb{/us save def}bdf
/pse{us restore}bdf
/ct 40 string def
/nc{currentpoint initclip newpath gc{dup type dup/arraytype eq
 exch/packedarraytype eq or{exec}if}
forall clip newpath moveto}def
/kp{ct 0 2 index length 2 index 39 2 index put getinterval copy cvx exec
 mx3 currentmatrix pop}bdf
end
LW 1 eq userdict/a4small known not and{/a4small
[[300 72 div 0 0 -300 72 div -120 3381]
280 3255
{statusdict/jobstate (printing) put 0 setblink
margins
exch 196 add exch 304 add 8 div round cvi frametoroket
statusdict/jobstate (busy) put
1 setblink}
/framedevice load
60 45{dup mul exch dup mul add 1.0 exch sub}/setscreen load
{}/settransfer load/initgraphics load/erasepage load]cvx
statusdict begin bind end readonly def}if
md begin/bigs[lnop userdict/letter known{/letter load}{lnop}ifelse
 userdict/legal known{/legal load}{lnop}ifelse userdict/a4 known{/a4
 load}{lnop}ifelse userdict/b5 known{/b5 load}{lnop}ifelse
lnop lnop lnop /tab load/a3Size load]def
/smalls[lnop userdict/lettersmall known{/lettersmall load}{userdict/note
 known{/note load}{lnop}ifelse}ifelse
userdict/legal known{/legal load}{lnop}ifelse userdict/a4small known
{/a4small load}{lnop}ifelse
userdict/b5 known{/b5 load}{userdict/note known{/note load}{lnop}ifelse
}ifelse lnop lnop lnop /tab load/a3Size load]def end
systemdict/currentpacking known{setpacking}if
{currentfile eexec} ( 
 checkload
373A767D4B7FD94FE5903B7014B1B8D3BED02632C855D56F458B118ACF3AF73FC4EF5E81
F5749042B5F9CF1016D093B75F250B7D8280B2EACE05A37037F7BDF6E12226D7D4E2DF2C
52FAFD5FD40FE72A0D3AC4BD485D8369D4C87636E920D1DAF222D92155A9CB1667E715F0
B82799B37CC8F5B32B74B39CF494536DC39C7EF04A7BCB29E2CEC79073CADCCFB23B4AA1
363F876F5121B618071B7B4EB1E5DE75FAA2368A3E5DB2B198623AFE92AE9484270FE7F5
7A850E88C0D3EEA156611C91D8E480D4370B025CCA6929A2BF40AD3D01B2CB7EE6DFB46E
12A830542337F7819B67F9765210F76DB06F34DA5B13A11759305C582E16D2B854939F6D
9121F2A4F285282F5DCD3D15896D121E3D6F5BE79E087451BB0ED233CDBEF090D3B4AC2D
C34B97E70C61D95FB072B8C12D2ABD843520949A39DCF99E2C1AA8FBCD025E47E0A82A8D
96E75BAF40F52AD402495BBD4DE0F356C8B14E764874E639C9F045A0D1908EC6456EB6C5
B8A6F826192F767EF2C55A21C58F5F9CC1F59247B55F2387828C7FE89D5E7D8484D1BC86
CB6673BDBE4FE17DD9BDE95224FE645136F41330BF155A4DDE1B0A32233BF471CE58FBC6
60DC7E641B0A0D30018454E2191C414A3011FF3FED1C0D88FE1FF9F75DCC456D09794722
6FBEC92509146D3A4CFFC0471B31C53222ED9DD88566F60F6C0D705AD79DACF53B070026
F083ED28B5CF757AAA0A169F6F320A75E9D2ED50ABD939AF85B6346C2ADB25D168F10508
E1516D194C635E6B187FADEA0829DBF0390C0F003F0265E215BC96CA3CC13D4A8E01570B
E193CA75A620728CD275ACF1986EFFB3A13419FE55EA7C4467B7E7EEDC1FC29C9F8C46A5
57D2CCDB914EF7B93E7530D555DFC2398AFC68CAD991F062EF85BAA1884EC166C7C5DF85
43666D8C41BE267D706BD1588F1F662F705CAE4D29DC38EF66BFAA89470D8A099B6F1B45
87F7B024412276106FCD3EB5AE17A5D1DF1781992DC40EA0A992F706F701304CEA9D9073
E7A74F1E687D81C3E5841D31CF86855BAAAD9B5D30317C75150A857C6B114735315CDD1A
EF36C26BBB0645499406DEE2F24B3B1C72FEC97C7BA31AA2CDAB25418BB1DC4C7E4757F1
D625087B0FD0300C03A65F2A72CE734925735277E034CDCF599129679F70CC8B66E03878
851DB75041F275E1E5761F3EC753BE1359CA364A22047AE4886217F9259FE19FF5B116E8
019B98B143114B313E8BEF87EC949D85C82E0812E6F50525E73890AF362CC8EE8A85F419
7E6AC18638EF12E56A808D439AF1BFD363F140314BF4E534485C42F1856688CC35288E8D
770120A420FB9F1FCF8AE8BD6D6156CC23E6C51119FE4DE1B68C9DF3487E9974BF9ED31F
8D3CE93FF101867319F2FF492D5D398B4F09A66F2F55BCAB34B99173B7EE89039D00DD21
A7B3A52E9F028F8301B5FC12D409412E064513BC579AAC498F577EA8ECD1FE3E42DC3CC3
20786C7B00194FEDF344402C33FC492D4BA86992B01683F440220FFE756BC88A94223D31
6078D69D33560E8EAB76B24CB7AA4320CF435593D76F624324ABE00B5587A4F283C725EA
24567133F25F472B5E2E4474DDB5A16AC5F2DF32350395D3E3892FE361F4D5C9A610C654
C9227614FBBAFF3356A90A2266E00F66234061075491571A65616211257F160000000000
000000000000000000000000000000000000000000000000000000
0000000000000000000000000000000000000000000000000000000000000000
0000000000000000000000000000000000000000000000000000000000000000
0000000000000000000000000000000000000000000000000000000000000000
0000000000000000000000000000000000000000000000000000000000000000
0000000000000000000000000000000000000000000000000000000000000000
0000000000000000000000000000000000000000000000000000000000000000
0000000000000000000000000000000000000000000000000000000000000000
cleartomark

{currentfile eexec} ( 
 checkload
F94E00EE41A71C59E5CAEED1EDBCF23D1DBA1EE99B9BB356492923BD8B1BA83A87CEB0E0
7377A31FD6241E814681118E17DC7CACE570399506E6E441B871B6043831BD03EFC11DBB
D8001EE2FF8CFBD485065D455A2E15AC36F1A84AD8789FA6461199C7CD14CB9FD64D4B06
452B7FC0A8FC263F70F1CCB893295D4DE70ADAB771C0F84396FA98C60B11DA02ABA15729
8DF0A23621853BEF167443A985ADC09BEFFD51CB4D29179E2B34609EF38A49DA61F4BFC2
56A3DE0732D7D29754A194857B9C9E9971227AA1DD0611FBB10E44E5FF66C062D9C24ED3
290529330BC317825E876929582DB0E39B9FC5EFD20CC1D4F94920EB9C534D0DA90DE70D
25BC7287319CF28602B3F46633C242CAFC8905E960317E3C2FA20AB8DB06ADBAF292FC7B
A2CA14EE65DF28B99CC11666B70AD33E8E1D57D63D4B89ECC615AE5747C1CA752C833D8D
6DE54CD4A0350B44310555CE3BD2C615ADD27B634CDB350AF3A432CE78AACD2909A5B586
F666CD87919A36DB1CBE86B3CE281DFD01CD7E1B8A18A4B415CECBFF79A5C4390A15EA77
D14D6BE12BAB5A8268C3F286D0590060647CABED674443CD258F11415E866AB330A25169
1B61F2422A61AFE59B6B4FBDCF85ED9BA0F8E483C034089E6877FF5923698D3A0DC0EED6
B9CFD32DF0839BC4EA5F6D1FCB6DD0920391E57E84745131D02D100179F4E0A68EC0A5FF
6680A6F463D038B04AF63FFA13D743B995A26A743C26D387209023C91DE43DF047A16F32
8AC9DDC08573B38BE9EA341EA16C78EC32F3A1B36B90D95A50610F4D050EC1C33497F3F3
A81A1B4C8BEF0BA84EE2FAA32DC112DAC490AF53E1749C4A0D866CAF7B893E52383B0D38
065C333FB122B700D7246F7EE87D942AE3DB5C1DD77E9E76C80CC5AD63D28DFED0E229CE
604673F78CD47F258FDF5BF3A3EAEC5C9BC8E482D8DBA9D268A35DA8C095A690679ED212
3E8B8F5E4826FA3B199EAA5D482D4B6AA86572E387CECEB7149C8947F41D6339328A748A
17F8C4AD3B0555F1E409450BA0C564F1F488BB5096EB003568D4D5EF6489897E27409547
D0EE4487D30184793B0F27BD265A64BDB3EA6761569DA955620C612E718677B77D6D81B9
99C6298877AFE0D1D6F6F358377A8BD2402F669C64B972B3A065EF7DD4BDEFFFE17E63DB
8898FA6E69166B710AAD6BA2EA9AF61E4B8C8701638D4D6E4DFFFC192AEF6BC027095C4C
72D748979675BA29FAF61E75343E14E61034602E5A79CD2519796ED6A9CC4EDEA46A9B59
D4A807E786B5EE46F25B0360BC8E7C12D723122CDEEF247C9776F4C99C8EBED6828AA197
44B5ADF0D07D95D98B3072372388D41B0FAB1CCE2775170679575ECDCA13B22A17FE9C66
05C3445F58F1A829512DAB6C528F83580C8AA53C35D605F626F5AD0B7FC1EA87D69A835E
3F53A1F450FB0AF42A5772F89D92A50D10F15BDBDA409F50C0B8AB93FE8A16D029DD8BB5
C480D1466735ED4D9CAF637E5ECD6C2ECB6BF3B3EFBEE7AB936D2C568E3009D156B87CAC
B1FB3A48A70BC91B2EC35CC9147FFB1A524E2B2F2E4E2C1B12F1C1C63768BB95CD62FEC0
1CBA79B9FA282DD4DF49990F27FF8EE4E2DDE2F0ACD83BC9D4BE0090192C7A799967EC4D
C2D63C0835E22D4C4B366D7FDCF3A05A4B53DF780F986EF25C79B665D5C00EFF7F17C0BB
6D544F9D83A7FDAC47D9C5683A656011374253C918FF6EA64749DD971B2300DD5320033E
01EC591F6318CCE94CE2B81C04322EC52B624E50643B52391CCD2AB56396A2AD8E2D3CA6
1B80D9D4CC363B2DF7863526958CDF3497E36648406C317E58EC563E7C26149A2A3C643A
DFB39A8DD92974C6D2A2A9D7B71CDF3FEBBF32BB02E7B45CF53AAEAD5E963A4AA4AF9A14
9A08A4EC303D5F2369977E93F54897EEAD31B06C5845D63F49D65F8E5573962241A57CCD
717CE6CA8C784A11192943616EA059B51BC38429E18D0121FCBB6FBD5D909B0D89E616C6
6DEF6A0F165A7030BD911A1B120468329CBB006C8D37720E531CF31E878CB4AAAC137633
675C3D546F5162487AB35F470C042BDEB945E0F2532BF92AA6FD53434440221ECD3533A7
AA89900CB19EFE2CD872DF8B7969AF0D3B72BF31DC5DD69CA6460966F61AB17CB5079640
98DBA3AF122EEC3128A9BAFE1034493F372B36BD1351205E9043A67C544402D8BCE24358
C8A5CE33867A00794CF7097D59C88279A11EE9C854E7E7AAE881F9828C569D208F5F3337
5F59E9A3818CFA38AAD0CBFBA32F9F44A8BB79DE4C40E3886457C16DA4A27953AA1E9947
2E35F2323F0BAA5E37DC28CBA46FEFB73B190016055ADD4D27615D748499A0E1C4B8C7EC
339C1C4D95A813A85918A8D01EEB485DDCDCEA6EA3F2C2A9D85C139CD90CCB352634F9AF
E836BCAC0C274E352BA2071B5269D5DE4CCDE3FF990CBA974980C7332AE1545A9C60D5D1
459D3AE95C1AC065733AF14FADB440A110DD539563B8D850CD0704C52F3F7CCCB53630D7
76560CBD22D8FF08F5B354487A171AEC15F5F54DE9CAB668BCAC573E788D92762EF63E76
087005F4AC2D02E0CAC173C11BE62ACE5DC4D3374F2F9746C9981E125FF9AB8CAE76D130
39E2C54DFD708E028A619EA1ED78E6B46F06DF0D0B74BBEDD8C190C7C0CEBDE8F7A4888C
C36575313478DD2CFE392E9BB7B2416955D44B7024A3BA43FBF37293B386D64746D77488
95411D243FAEC50638F2AA33337D7FA018ADDAC5835A0DDFAE99AD6299DFB4CA6872C598
53E3AC12FC9E3D26629C5B49CF844C87B3C4BFBE3074E3A1CE6984758C20C661084381CD
6B4582D84F19C0000B5FC0DCB42B567E396031601C095D7016283EBE5F13CD8A3A374A74
DDBBABD36081149F8BC242085F2F7297CC97FD3B8BAD206D8AC9707A39ECCC7963B522E0
8DA391A1EF12DD4D746DBDDDCC0834F88160CF189A9645567CEC2F023A571AF0DFD15DB8
5B744C28C000DF53B05F8F210841F6E87A04F20C777B7C0BE6182BE2E90226E5301A1253
2A745F2FAAA81637CF11B78CD2B99A4D18B862D6C5DBD31793FB16A2D9AAD376D4484D75
AA833D0068B1D34DB74E3302480854E3B5484D8A47E39A89A2FA927BC3641EA7F8E004FD
E4C2F08D40D99F1ACB47CAF6887629BF6DFE12968D297596D28CE0CF148B12E7DCB49FB9
4F5ADBD214C3A6CE1E249831BA9EB8A189F2CE1ABE39A7B537253E369A508A2AF2ADB946
3F9B56BBBFF31D535FF997F537C6675C196E7ECBD493F652FA7CC6D9C1CA3379BFDB5AF7
513C6E834054494296B91A6EE800114363D5D5D0759F41B4DECB653B9DE3E94583579EF5
49ED5F3FAFB12661ABC0C57A332406517ED3454EDED34B386C60F78DC976266E0EAF54FC
245FB0E3EFC8016236436B599C1C97A8C5E0AC8F7836161873C71F01ED9CC25C236420F4
1FD8277993D3959205912FA0927B59E3DAE7377D82079447D6E41EE5AEC0DFFF79AF8F4E
D47F17EE708FEA45877860D56F8CBCE65A061E8E1CA4A5FBAF0E13429A7F0ADB6F178FA4
49F46CC539BBC0107E3A53B1C362A04B20E6D721E7E6E1E4976A11DDC98C7614D22B53DF
BB6DAE533AC9BE882021A735C30DAA4A44AED09F49A390E8CFF59BD9C30667AF21B03EC5
CEBD5C2C3AA2769E8D714191A48E7DDF50B13D1560E82EFB65FCE601AE9E8C351FBA1DED
80B7351314E7F9F9A784BFE3759B7E322A84E7B51F9DC5F5D9C8050CD79B27C0A4B0DD68
A3C27A948AD6858E35B960D2DEA838C479CAEA83B1A912174ACB2100E55E7A14892D7A9B
3711FF0B20065C1995B49E1F23464A92DD140642E3A7B1973849E64D1A3CF60000000000
000000000000000000000000000000000000000000000000000000
0000000000000000000000000000000000000000000000000000000000000000
0000000000000000000000000000000000000000000000000000000000000000
0000000000000000000000000000000000000000000000000000000000000000
0000000000000000000000000000000000000000000000000000000000000000
0000000000000000000000000000000000000000000000000000000000000000
0000000000000000000000000000000000000000000000000000000000000000
0000000000000000000000000000000000000000000000000000000000000000
cleartomark


md begin
F sgd
svsc

T T 0 0 3040 2300 -130 -125 3170 2425 100 300 300 1 F F F F T T T F psu
(Norman-B32613A-P049064; document: plotls)jn
0 mf
od
op
0 0 3040 2300 fr
0 0 xl
1 1 pen
0 0 gm
(nc 0 0 0 0 6 rc)kp
1496 610 gm
(nc 979 610 2181 1810 6 rc)kp
1499 616 lin
1502 622 lin
1504 628 lin
1507 634 lin
1509 640 lin
1512 646 lin
1514 652 lin
1516 658 lin
1518 664 lin
1520 670 lin
1522 676 lin
1524 682 lin
1526 688 lin
1528 694 lin
1530 700 lin
1531 706 lin
1533 712 lin
1535 718 lin
1537 724 lin
1538 730 lin
1539 736 lin
1541 742 lin
1542 748 lin
1544 754 lin
1545 760 lin
1546 766 lin
1548 772 lin
1549 778 lin
1550 784 lin
1552 790 lin
1553 796 lin
1554 802 lin
1555 808 lin
1556 814 lin
1557 820 lin
1558 826 lin
1559 832 lin
1561 838 lin
1562 844 lin
1562 850 lin
1563 856 lin
1564 862 lin
1565 868 lin
1566 874 lin
1567 880 lin
1568 886 lin
1569 892 lin
1570 898 lin
1571 904 lin
1572 910 lin
1573 916 lin
1574 922 lin
1575 928 lin
1576 934 lin
1577 940 lin
1577 946 lin
1578 952 lin
1579 958 lin
1580 964 lin
1581 970 lin
1582 976 lin
1582 982 lin
1583 988 lin
1584 994 lin
1585 1000 lin
1586 1006 lin
1587 1012 lin
1587 1018 lin
1588 1024 lin
1589 1030 lin
1590 1036 lin
1591 1042 lin
1592 1048 lin
1593 1054 lin
1593 1060 lin
1594 1066 lin
1595 1072 lin
1596 1078 lin
1597 1084 lin
1598 1090 lin
1599 1096 lin
1599 1102 lin
1600 1108 lin
1601 1114 lin
1602 1120 lin
1603 1126 lin
1604 1132 lin
1605 1138 lin
1606 1144 lin
1607 1150 lin
1608 1156 lin
1609 1162 lin
1610 1168 lin
1611 1174 lin
1612 1180 lin
1613 1186 lin
1614 1192 lin
1615 1198 lin
1616 1204 lin
1617 1210 lin
1619 1216 lin
1620 1222 lin
1621 1228 lin
1622 1234 lin
1623 1240 lin
1625 1246 lin
1626 1252 lin
1627 1258 lin
1628 1264 lin
1630 1270 lin
1631 1276 lin
1633 1282 lin
1634 1288 lin
1636 1294 lin
1637 1300 lin
1639 1306 lin
1641 1312 lin
1642 1318 lin
1644 1324 lin
1646 1330 lin
1648 1336 lin
1650 1342 lin
1652 1348 lin
1653 1354 lin
1656 1360 lin
1658 1366 lin
1660 1372 lin
1662 1378 lin
1665 1384 lin
1667 1390 lin
1670 1396 lin
1672 1402 lin
1675 1408 lin
1678 1414 lin
1681 1420 lin
1684 1426 lin
1687 1432 lin
1691 1438 lin
1695 1444 lin
1698 1450 lin
1702 1456 lin
1706 1462 lin
1711 1468 lin
1715 1474 lin
1720 1480 lin
1725 1486 lin
1730 1492 lin
1736 1498 lin
1742 1504 lin
1748 1510 lin
1754 1516 lin
1761 1522 lin
1769 1528 lin
1777 1534 lin
1786 1540 lin
1795 1546 lin
1805 1552 lin
1816 1558 lin
1828 1564 lin
1840 1570 lin
1854 1576 lin
1869 1582 lin
1886 1588 lin
1904 1594 lin
1924 1600 lin
1947 1606 lin
1972 1612 lin
2002 1618 lin
2034 1624 lin
2072 1630 lin
2117 1636 lin
2168 1642 lin
2180 1643 lin
1457 610 gm
1460 616 lin
1462 622 lin
1465 628 lin
1467 634 lin
1470 640 lin
1472 646 lin
1474 652 lin
1476 658 lin
1478 664 lin
1480 670 lin
1482 676 lin
1484 682 lin
1485 688 lin
1487 694 lin
1488 700 lin
1490 706 lin
1491 712 lin
1493 718 lin
1494 724 lin
1496 730 lin
1497 736 lin
1498 742 lin
1500 748 lin
1501 754 lin
1502 760 lin
1503 766 lin
1504 772 lin
1505 778 lin
1506 784 lin
1507 790 lin
1508 796 lin
1509 802 lin
1510 808 lin
1511 814 lin
1512 820 lin
1513 826 lin
1514 832 lin
1515 838 lin
1515 844 lin
1516 850 lin
1517 856 lin
1518 862 lin
1519 868 lin
1519 874 lin
1520 880 lin
1521 886 lin
1521 892 lin
1522 898 lin
1523 904 lin
1523 910 lin
1524 916 lin
1525 922 lin
1525 928 lin
1526 934 lin
1527 940 lin
1527 946 lin
1528 952 lin
1528 958 lin
1529 964 lin
1530 970 lin
1530 976 lin
1531 982 lin
1531 988 lin
1532 994 lin
1532 1000 lin
1533 1006 lin
1534 1012 lin
1534 1018 lin
1535 1024 lin
1535 1030 lin
1536 1036 lin
1536 1042 lin
1537 1048 lin
1537 1054 lin
1538 1060 lin
1538 1066 lin
1539 1072 lin
1540 1078 lin
1540 1084 lin
1541 1090 lin
1541 1096 lin
1542 1102 lin
1542 1108 lin
1543 1114 lin
1544 1120 lin
1544 1126 lin
1545 1132 lin
1545 1138 lin
1546 1144 lin
1546 1150 lin
1547 1156 lin
1548 1162 lin
1548 1168 lin
1549 1174 lin
1550 1180 lin
1550 1186 lin
1551 1192 lin
1552 1198 lin
1553 1204 lin
1553 1210 lin
1554 1216 lin
1555 1222 lin
1555 1228 lin
1556 1234 lin
1557 1240 lin
1558 1246 lin
1559 1252 lin
1560 1258 lin
1561 1264 lin
1562 1270 lin
1563 1276 lin
1564 1282 lin
1565 1288 lin
1566 1294 lin
1567 1300 lin
1568 1306 lin
1569 1312 lin
1570 1318 lin
1572 1324 lin
1573 1330 lin
1574 1336 lin
1576 1342 lin
1577 1348 lin
1579 1354 lin
1580 1360 lin
1582 1366 lin
1584 1372 lin
1586 1378 lin
1587 1384 lin
1589 1390 lin
1592 1396 lin
1594 1402 lin
1596 1408 lin
1598 1414 lin
1601 1420 lin
1603 1426 lin
1606 1432 lin
1609 1438 lin
1612 1444 lin
1615 1450 lin
1618 1456 lin
1622 1462 lin
1625 1468 lin
1629 1474 lin
1634 1480 lin
1638 1486 lin
1643 1492 lin
1647 1498 lin
1653 1504 lin
1658 1510 lin
1664 1516 lin
1671 1522 lin
1678 1528 lin
1685 1534 lin
1693 1540 lin
1702 1546 lin
1711 1552 lin
1721 1558 lin
1732 1564 lin
1744 1570 lin
1757 1576 lin
1771 1582 lin
1787 1588 lin
1805 1594 lin
1824 1600 lin
1846 1606 lin
1871 1612 lin
1899 1618 lin
1931 1624 lin
1968 1630 lin
2012 1636 lin
2063 1642 lin
2124 1648 lin
2180 1652 lin
1314 610 gm
1317 616 lin
1319 622 lin
1321 628 lin
1323 634 lin
1325 640 lin
1327 646 lin
1329 652 lin
1331 658 lin
1332 664 lin
1334 670 lin
1336 676 lin
1337 682 lin
1338 688 lin
1340 694 lin
1341 700 lin
1342 706 lin
1343 712 lin
1345 718 lin
1345 724 lin
1346 730 lin
1347 736 lin
1348 742 lin
1349 748 lin
1350 754 lin
1351 760 lin
1352 766 lin
1353 772 lin
1353 778 lin
1354 784 lin
1354 790 lin
1355 796 lin
1356 802 lin
1356 808 lin
1357 814 lin
1357 820 lin
1358 826 lin
1358 832 lin
1359 838 lin
1359 844 lin
1359 850 lin
1360 856 lin
1360 862 lin
1361 868 lin
1361 874 lin
1361 880 lin
1361 886 lin
1362 892 lin
1362 898 lin
1362 904 lin
1362 910 lin
1362 916 lin
1362 922 lin
1363 928 lin
1363 934 lin
1363 940 lin
1363 946 lin
1363 952 lin
1363 958 lin
1363 964 lin
1363 970 lin
1363 976 lin
1363 982 lin
1363 988 lin
1363 994 lin
1363 1000 lin
1363 1006 lin
1363 1012 lin
1363 1018 lin
1363 1024 lin
1363 1030 lin
1363 1036 lin
1363 1042 lin
1363 1048 lin
1363 1054 lin
1363 1060 lin
1363 1066 lin
1363 1072 lin
1363 1078 lin
1363 1084 lin
1363 1090 lin
1363 1096 lin
1363 1102 lin
1363 1108 lin
1363 1114 lin
1363 1120 lin
1363 1126 lin
1363 1132 lin
1362 1138 lin
1362 1144 lin
1362 1150 lin
1362 1156 lin
1362 1162 lin
1362 1168 lin
1362 1174 lin
1362 1180 lin
1362 1186 lin
1362 1192 lin
1362 1198 lin
1362 1204 lin
1362 1210 lin
1362 1216 lin
1362 1222 lin
1362 1228 lin
1362 1234 lin
1362 1240 lin
1362 1246 lin
1362 1252 lin
1362 1258 lin
1363 1264 lin
1363 1270 lin
1363 1276 lin
1363 1282 lin
1363 1288 lin
1364 1294 lin
1364 1300 lin
1364 1306 lin
1365 1312 lin
1365 1318 lin
1365 1324 lin
1366 1330 lin
1366 1336 lin
1367 1342 lin
1368 1348 lin
1368 1354 lin
1369 1360 lin
1370 1366 lin
1370 1372 lin
1371 1378 lin
1372 1384 lin
1373 1390 lin
1374 1396 lin
1375 1402 lin
1377 1408 lin
1378 1414 lin
1379 1420 lin
1381 1426 lin
1383 1432 lin
1385 1438 lin
1387 1444 lin
1388 1450 lin
1391 1456 lin
1393 1462 lin
1396 1468 lin
1399 1474 lin
1402 1480 lin
1405 1486 lin
1409 1492 lin
1412 1498 lin
1416 1504 lin
1421 1510 lin
1426 1516 lin
1431 1522 lin
1437 1528 lin
1443 1534 lin
1450 1540 lin
1457 1546 lin
1465 1552 lin
1474 1558 lin
1484 1564 lin
1494 1570 lin
1506 1576 lin
1519 1582 lin
1534 1588 lin
1550 1594 lin
1568 1600 lin
1589 1606 lin
1612 1612 lin
1639 1618 lin
1670 1624 lin
1706 1630 lin
1748 1636 lin
1797 1642 lin
1857 1648 lin
1931 1654 lin
2023 1660 lin
2141 1666 lin
2180 1667 lin
1319 610 gm
1321 616 lin
1323 622 lin
1324 628 lin
1326 634 lin
1327 640 lin
1329 646 lin
1329 652 lin
1331 658 lin
1332 664 lin
1333 670 lin
1333 676 lin
1334 682 lin
1335 688 lin
1336 694 lin
1336 700 lin
1337 706 lin
1337 712 lin
1337 718 lin
1337 724 lin
1338 730 lin
1338 736 lin
1338 742 lin
1338 748 lin
1338 754 lin
1338 760 lin
1338 766 lin
1338 772 lin
1338 778 lin
1338 784 lin
1337 790 lin
1337 796 lin
1337 802 lin
1337 808 lin
1336 814 lin
1336 820 lin
1335 826 lin
1335 832 lin
1334 838 lin
1333 844 lin
1333 850 lin
1332 856 lin
1331 862 lin
1330 868 lin
1330 874 lin
1329 880 lin
1328 886 lin
1327 892 lin
1326 898 lin
1325 904 lin
1324 910 lin
1323 916 lin
1322 922 lin
1321 928 lin
1320 934 lin
1318 940 lin
1317 946 lin
1316 952 lin
1314 958 lin
1313 964 lin
1312 970 lin
1310 976 lin
1309 982 lin
1307 988 lin
1306 994 lin
1304 1000 lin
1303 1006 lin
1301 1012 lin
1299 1018 lin
1297 1024 lin
1296 1030 lin
1294 1036 lin
1292 1042 lin
1290 1048 lin
1288 1054 lin
1286 1060 lin
1284 1066 lin
1282 1072 lin
1280 1078 lin
1278 1084 lin
1275 1090 lin
1273 1096 lin
1271 1102 lin
1268 1108 lin
1266 1114 lin
1263 1120 lin
1261 1126 lin
1258 1132 lin
1255 1138 lin
1253 1144 lin
1250 1150 lin
1247 1156 lin
1244 1162 lin
1241 1168 lin
1238 1174 lin
1235 1180 lin
1232 1186 lin
1228 1192 lin
1225 1198 lin
1221 1204 lin
1218 1210 lin
1214 1216 lin
1211 1222 lin
1207 1228 lin
1203 1234 lin
1199 1240 lin
1195 1246 lin
1191 1252 lin
1187 1258 lin
1182 1264 lin
1178 1270 lin
1173 1276 lin
1168 1282 lin
1163 1288 lin
1158 1294 lin
1153 1300 lin
1148 1306 lin
1142 1312 lin
1137 1318 lin
1131 1324 lin
1125 1330 lin
1119 1336 lin
1113 1342 lin
1106 1348 lin
1099 1354 lin
1093 1360 lin
1085 1366 lin
1078 1372 lin
1070 1378 lin
1063 1384 lin
1054 1390 lin
1046 1396 lin
1037 1402 lin
1028 1408 lin
1018 1414 lin
1008 1420 lin
998 1426 lin
988 1432 lin
979 1436 lin
1580 610 gm
1580 621 lin
1580 641 gm
1580 652 lin
1580 671 gm
1580 683 lin
1580 702 gm
1580 714 lin
1580 733 gm
1580 745 lin
1580 764 gm
1580 775 lin
1580 795 gm
1580 806 lin
1580 825 gm
1580 837 lin
1580 856 gm
1580 868 lin
1580 887 gm
1580 898 lin
1580 918 gm
1580 929 lin
1580 948 gm
1580 960 lin
1580 979 gm
1580 991 lin
1580 1010 gm
1580 1021 lin
1580 1041 gm
1580 1052 lin
1580 1071 gm
1580 1083 lin
1580 1102 gm
1580 1114 lin
1580 1133 gm
1580 1145 lin
1580 1164 gm
1580 1175 lin
1580 1195 gm
1580 1206 lin
1580 1225 gm
1580 1237 lin
1580 1256 gm
1580 1268 lin
1580 1287 gm
1580 1298 lin
1580 1318 gm
1580 1329 lin
1580 1348 gm
1580 1360 lin
1580 1379 gm
1580 1391 lin
1580 1410 gm
1580 1421 lin
1580 1441 gm
1580 1452 lin
1580 1471 gm
1580 1483 lin
1580 1502 gm
1580 1514 lin
1580 1533 gm
1580 1545 lin
1580 1564 gm
1580 1575 lin
1580 1595 gm
1580 1606 lin
1580 1625 gm
1580 1637 lin
1580 1656 gm
1580 1668 lin
1580 1687 gm
1580 1698 lin
1580 1718 gm
1580 1729 lin
1580 1750 gm
1580 1762 lin
1580 1782 gm
1580 1794 lin
1457 610 gm
1457 621 lin
1457 641 gm
1457 652 lin
1457 671 gm
1457 683 lin
1457 702 gm
1457 714 lin
1457 733 gm
1457 745 lin
1457 764 gm
1457 775 lin
1457 795 gm
1457 806 lin
1457 825 gm
1457 837 lin
1457 856 gm
1457 868 lin
1457 887 gm
1457 898 lin
1457 918 gm
1457 929 lin
1457 948 gm
1457 960 lin
1457 979 gm
1457 991 lin
1457 1010 gm
1457 1021 lin
1457 1041 gm
1457 1052 lin
1457 1071 gm
1457 1083 lin
1457 1102 gm
1457 1114 lin
1457 1133 gm
1457 1145 lin
1457 1164 gm
1457 1175 lin
1457 1195 gm
1457 1206 lin
1457 1225 gm
1457 1237 lin
1457 1256 gm
1457 1268 lin
1457 1287 gm
1457 1298 lin
1457 1318 gm
1457 1329 lin
1457 1348 gm
1457 1360 lin
1457 1379 gm
1457 1391 lin
1457 1410 gm
1457 1421 lin
1457 1441 gm
1457 1452 lin
1457 1471 gm
1457 1483 lin
1457 1502 gm
1457 1514 lin
1457 1533 gm
1457 1545 lin
1457 1564 gm
1457 1575 lin
1457 1595 gm
1457 1606 lin
1457 1625 gm
1457 1637 lin
1457 1656 gm
1457 1668 lin
1457 1687 gm
1457 1698 lin
1457 1718 gm
1457 1729 lin
1457 1750 gm
1457 1762 lin
1457 1782 gm
1457 1794 lin
2181 610 gm
(nc 618 250 2422 2050 6 rc)kp
2181 585 lin
1980 610 gm
1980 585 lin
1780 610 gm
1780 585 lin
1580 610 gm
1580 585 lin
1379 610 gm
1379 585 lin
1179 610 gm
1179 585 lin
979 610 gm
979 585 lin
2181 1810 gm
2181 1835 lin
1980 1810 gm
1980 1835 lin
1780 1810 gm
1780 1835 lin
1580 1810 gm
1580 1835 lin
1379 1810 gm
1379 1835 lin
1179 1810 gm
1179 1835 lin
979 1810 gm
979 1835 lin
2181 610 gm
2206 610 lin
2181 910 gm
2206 910 lin
2181 1210 gm
2206 1210 lin
2181 1510 gm
2206 1510 lin
2181 1810 gm
2206 1810 lin
979 610 gm
954 610 lin
979 910 gm
954 910 lin
979 1210 gm
954 1210 lin
979 1510 gm
954 1510 lin
979 1810 gm
954 1810 lin
979 610 gm
2181 610 lin
2181 1810 lin
979 1810 lin
979 610 lin
8 8 pen
622 254 gm
64 gr
622 254 lin
3040 2300 12665 9583 th
2210 504 gm
F 1 setTxMode
0 fs
bu fc
{}mark T /Helvetica /|______Helvetica 0 rf
bn
12.47991 fz
bu fc
2 F /|______Helvetica fnt
bn
3.90951 0.(-6)ashow
tu
ts
2010 504 gm
3.90951 0.(-4)ashow
tu
ts
1810 504 gm
3.90951 0.(-2)ashow
tu
ts
1610 533 gm
(0)show
tu
ts
1410 533 gm
(2)show
tu
ts
1205 533 gm
(4)show
tu
ts
1005 533 gm
(6)show
tu
ts
2310 554 gm
0.90658 0.(0.00)ashow
tu
nc ct 39 0 put
5 0 270 tp
0 0 0 0 tc
ts
1660 371 gm
tt
2.42752 0. 32 0.24275 0.(V \(eV\))awidthshow
te
tu
(nc 618 250 2422 2050 6 rc)kp
ts
2310 854 gm
0.90658 0.(0.05)ashow
tu
ts
2310 1154 gm
0.90658 0.(0.10)ashow
tu
ts
2310 1454 gm
0.90658 0.(0.15)ashow
tu
ts
2310 1754 gm
0.90658 0.(0.20)ashow
tu
ts
2380 1092 gm
bu fc
{}mark F /Symbol /|______Symbol 0 rf
bn
12 fz
bu fc
2 F /|______Symbol fnt
bn
(c)show
2410 1121 gm
9.35993 fz
bu fc
2 F /|______Helvetica fnt
bn
(0)show
2380 1146 gm
12.47991 fz
bu fc
2 F /|______Helvetica fnt
bn
( )show
2380 1163 gm
1.40553 0.(\(1/eV\))ashow
tu
ts
1114 1221 gm
9.35993 fz
bu fc
2 F /|______Helvetica fnt
bn
(1)show
1143 1246 gm
12.47991 fz
bu fc
2 F /|______Helvetica fnt
bn
(I)show
tu
ts
1314 1350 gm
9.35993 fz
bu fc
2 F /|______Helvetica fnt
bn
(3)show
1343 1375 gm
12.47991 fz
bu fc
2 F /|______Helvetica fnt
bn
(P)show
tu
ts
1522 1296 gm
9.35993 fz
bu fc
2 F /|______Helvetica fnt
bn
(3)show
1551 1321 gm
12.47991 fz
bu fc
2 F /|______Helvetica fnt
bn
(F)show
tu
ts
1710 1254 gm
9.35993 fz
bu fc
2 F /|______Helvetica fnt
bn
(3)show
1739 1279 gm
12.47991 fz
bu fc
2 F /|______Helvetica fnt
bn
(H)show
tu
ts
1614 1858 gm
(f)show
1585 1883 gm
9.35993 fz
bu fc
2 F /|______Helvetica fnt
bn
(1)show
tu
ts
1485 1850 gm
12.47991 fz
bu fc
2 F /|______Helvetica fnt
bn
(f)show
1455 1875 gm
9.35993 fz
bu fc
2 F /|______Helvetica fnt
bn
(2)show
tu
F T cp
cd
end